\documentclass[secnumarabic,amssymb, amsmath, nofootinbib,
nobibnotes, aps, prl]{revtex4}
\usepackage{dcolumn}
\usepackage{longtable}
\usepackage{graphicx}
\begin{document}
\title[Electron impact collision strengths ...]{Electron impact collision strengths in Si IX,
Si X and Si XI}
\author{Guiyun Liang$^{a}$}
\author{Gang Zhao$^a$}\email{gzhao@bao.ac.cn}
\author{Jiaolong Zeng$^{b}$}
\affiliation{$^a$National Astronomical Observatories, Chinese
Academy of Sciences, Beijing 100012, P. R. China}
\affiliation{$^b$Department of Applied Physics, National
University of Defense Technology, Changsha 410073, P. R. China}

\begin{abstract} Electron impact collision
strengths among 560 levels of Si IX, 320 levels of Si X and 350
levels of Si XI have been calculated using the Flexible Atomic
Code (FAC) of Gu (2003). Collision strengths $\Omega$ at ten
scattered electron energies covering an entire energy range,
namely 10, 50, 100, 200, 400, 600, 800, 1000, 1500 and 2000~eV,
are reported. Assuming a Maxwellian energy distribution, effective
collision strengths $\Upsilon$ are obtained at a finer electron
temperature grids of 0.5, 1.0, 2.0, 3.0, 4.0, 5.0 and 6.0~MK,
which covers the typical temperature range of astrophysical hot
plasma. Additionally, radiative rates $A$ and weighted oscillator
strengths $gf$ are given for more possible transitions among these
levels. Comparisons of our results with available predictions
reported in earlier literatures are made, and the accuracy of the
data is assessed. Most transitions exhibit a better agreement,
whereas large differences in $gf$ appear for a few cases, which
are due to the different inclusion of configuration interaction in
different theoretical calculations. In excitations among levels of
ground and lower excited configurations, large discrepancies of
$\Upsilon$ maybe resulted from the consideration of resonance
effects in earlier works.
\end{abstract}
\maketitle

\section{1. Introduction}
A wealth of high resolution spectra in UV, EUV and X-ray regions
has been obtained for solar, stellar and other astrophysical
sources by many space missions, such as {\it SOHO}, {\it Chandra}
and {\it XMM-Newton}. Many of the observed emission lines are due
to highly charged silicon ions (Si VII--Si XIV), as reported in
literatures for Procyon and $\alpha$ Centauri~\cite{RMA02,RNM03}.
For line identifications and spectral analyses, a complete list of
lines including emission or absorption lines of highly charged
silicon, is very necessary. This is available in the
Chianti~\footnotemark[1]\footnotetext[1]{http://wwwsolar.nrl.navy.mil/chianti.html}
and
ATOMDB~\footnotemark[2]\footnotetext[2]{http://cxc.harvard.edu/atomdb/}
data sets. However, the available data of atomic parameters is
very limited for elements such as silicon, sulfur, argon and
calcium. Moreover, the accuracy of the available data for these
L-shell ions attracts attention because of the poor modelling for
the astrophysical spectra~\cite{ABG01}. Recently, Aggarwal et
al.~\cite{AKN05}, Liang et al.~\cite{LDZ04} and Landi \&
Bhatia~\cite{LB06} performed the atomic data calculations of Ar
L-shell ions with larger configuration interaction~(CI). They
listed energy levels, radiative rates, and/or collision strengths.
Generally, more accurate results could be obtained by considering
larger CI, as stated by Aggarwal et al.~\cite{AKN05}.

Although many calculations for the highly charged silicon ions
have been performed in past decades, and the data is used
extensively in present astrophysical modelling codes, such as
Chianti and APEC. Yet the data is limited to low-lying energy
levels. For electron impact collision strengths, almost all
available data is confined to levels with $n=2$ configurations.
Moreover, available theoretical data of radiative rates is also
confined to allowed and inter-combination (E1) transitions alone.
In addition, the data used by the Chianti and APEC codes is from
different literatures. Here, we attempt to present a self-consist,
accurate and extensive data for highly charged Si ions.
Additionally, to provide experimental support, many lines of Si
VIII
--- XIII in the 40---80~\AA\, soft X-ray range have been measured
in an experiment of silicon target irradiated by intense
femtosecond laser at the Institute of Physics, Chinese Academy of
Sciences~\cite{LZJ06}.

\section{2. Energy levels}
This study adopts the flexible atomic code (FAC) developed by
Gu~\cite{Gu03,Gu05} to perform calculations of the structure and
the $e$-ion interaction process. This code is a standard atomic
structure code alike GRASP code of Dyall et al.~\cite{DGJ89}, and
available at the website
\url{http://kipac-tree.stanford.edu/fac/}. A fully relativistic
approach based on Dirac equation is used throughout the entire
package. The specific configurations included, results obtained
and accuracy achieved, are discussed below for each ion.

\subsection{2.1 Si IX}
In the work of Aggarwal~\cite{Agg98}, 46 low-lying energy levels
and radiative rates among these levels for carbon-like ions
including Si IX, were reported. An extensive CI was considered in
this work, and better results were obtained when compared with
earlier calculations. Yet only 46 levels belong to configurations
of ($1s^2$)$2s^22p^2$, $2s2p^3$, $2p^4$, and $2s^22p3l$, were
reported. In comparison with experimental values
(NIST\footnotemark[1]\footnotetext[1]{http://physics.nist.gov/cgi-bin/AtData/main$\_$asd}),
level energies agree within 2\% except for two levels of
$2s^22p^2~^1D_2$ and $^1S_0$. The Chianti code adopts an earlier
calculation of Bhatia \& Doschek~\cite{BD93}, which also reported
46 levels, but less CI was considered. Besides atomic data such as
the energy levels, radiative decay rates and collision strengths,
this work also presented line intensities for some strong
transitions by solving rate equations of level populations.
Orloski et al.~\cite{OTC99} (hereafter OTC99) performed the
calculation of energy levels by including additional
configurations, namely $2s2p^23l$. Further, the calculated energy
levels were adjusted again by observed wavelengths using an
interactive optimization procedure packaged in the program
ELCALC~\cite{RK69}.

In our study, energies of 560 levels belonging to 31
configurations of Si IX [namely, $2s^22p^2$, $2s2p^3$, $2p^4$,
$2s^22p3l$, $2s2p^23l$, $2p^33l$ ($l=s, p, d$), $2s^22p4l'$,
$2s2p^24l'$, $2s^22p5l'$, $2s^22p6l'$ ($l'=0, 1, ..., n-1$)] are
reported as listed in Table 1. In order to assess the accuracy, we
also list the NIST data which is recognized the most reliable data
so far, and other theoretical calculations. In comparison with
available experimental values, present results are better than 2\%
for most levels. Though a less CI has been considered than the
work of Aggarwal~\cite{Agg98}, present level energies show an
excellent agreement with those reported by Aggarwal~\cite{Agg98}.
On the other hand, this indicates that the CI effect from other
configurations is not distinct. The theoretical calculation of
Bhatia \& Doschek~\cite{BD93} also agree with experimental values
except for the level $2s2p^3~^5S_2$. Its value is lower than the
experimental one by $\sim$ 8\%. By considering large CI, the level
energy increases to 1.3025~{\it Ryd}, and agrees with experimental
one within 2\%. Results of Orloski et al.~\cite{OTC99} had been
adjusted by observed wavelengths, so their level energies are
listed for comparison in Table 1. Fig.1-(a) visually illustrate
such comparison. For those higher excited levels, the
configuration mixing is very severe. So level designation and
ordering are very difficult. And only a few levels can be obtained
from NIST database, and some levels are labelled by question mark
`?'. For these levels, we note that our results differ the
experimental values by less than 0.07$Ryd$. This suggest that a
good agreement has been obtained. Aggarwal et al.~\cite{Agg98}
considered a larger CI in their calculation. Unfortunately, the
higher excited levels have not been reported.
\begin{figure}
\includegraphics[angle=-90,width=9cm]{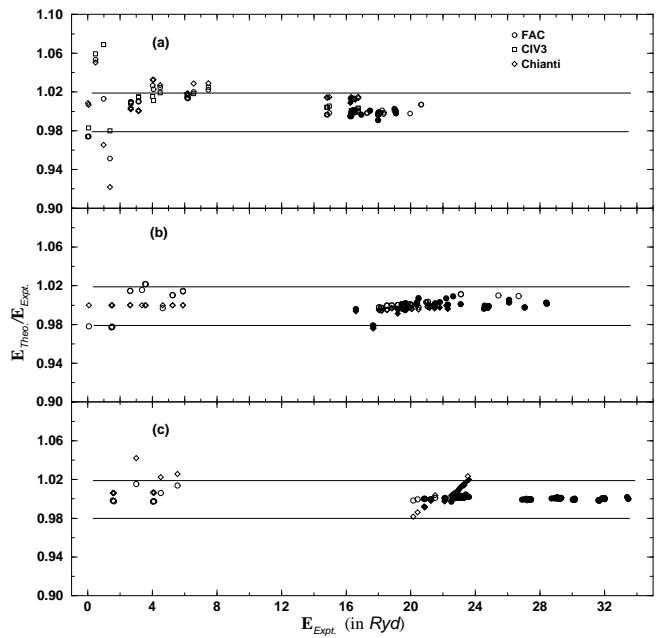}
\caption{Comparison of different calculations of energy levels
versus the available experimental ones (from NIST database). The
x-axis denotes the experimental energy (in $Ryd$), the y-axis
denotes ratio of the theoretical calculations {\it vs}
experimental ones. (a) for Si IX; (b) for Si X; (c) for Si XI.}
\end{figure}

\subsection{2.2 Si X}
For this ion, 50 experimental level energies belonging to the
$2s^22p$, $2s2p^2$, $2p^3$, $2s2p3l~(l =s, p, d)$, $2p^23d$,
$2s^24d$, $2s^25d$ and $2s^26d$ configurations can be obtained
from the NIST website. In the Chianti database, additional 75
theoretical energy levels belonging to configurations, namely
$2s^22p$, $2s2p^2$, $2p^3$, $2s^23l$, $2s2p3l$, $2p^23l~(l =s, p,
d)$, are compiled from earlier works. And different theoretical
calculations from different atomic physicists have been included,
such as Edlen~\cite{Edl83}, Zhang et al.~\cite{ZS94,ZGP94}, and
Sampson \& Zhang~\cite{SZ95}. Cavalcanti et al.~\cite{CLT00} also
reported some level energies belonging to $2s^22p$, $2p^3$,
$2s^23l$, $2s2p3l$, $2p^23l$, $2s^24l$, $2s2p4p$, $2s2p4d$,
$2s^25d$, $2s2p5p$ and $2s2p5d$ configurations using a
multi-configuration Hartree-Fock relativistic approach, and
adjusted their results by observed wavelengths.

By considered a larger CI, we calculate the level energies of
Si~X, and 320 levels belonging to configurations of $2s^22p$,
$2p^3$, $2s^23l$, $2s2p3l$, $2p^23l~(l =s, p, d)$, $2s^24l'$,
$2s2p4l'$, $2s^25l'$, $2s2p5l'$ and $2s^26l'$~($l'= 0, 1, ...,
n-1$; $n$ is main quantum number) are reported as listed in Table
2. For comparison, other available theoretical results and
experimental ones are also presented.

An inspection of Table 2 indicates that present results show a
good agreement with experimental values and other different
predictions, which is visually displayed in Fig.1-(b). Some higher
excited levels such as levels from the $2s2p4l$ and $2s2p5l$
configurations, were also reported by Cavalcanti et
al.~\cite{CLT00} (hereafter referring to CLT00), and these energy
values were adjusted by observed wavelengths.  A comparison with
these results reveals that our results are better than 1\% as
shown by filled circle in Fig.1-(b). From Table 2, we found the
differences are less than 0.08~$Ryd$.

\subsection{2.3 Si XI}
For this ion, almost all available theoretical results of energy
levels are for low-lying 46 levels with $n=3$ configurations,
which are extensively used by current astrophysical modelling
codes such as Chianti, MEKAL and APEC. Typical studies are the
works of Sampson et al.~\cite{SG84} and Zhang $\&$
Sampson~\cite{ZS92}. Coutinho \& Trigueiros~\cite{CT01} also
reported higher excited levels of $2s4l, 2p4l, 2s5l$ and $2p5l$
configurations, in which a multi-configuration Hartree-Fock
relativistic approach was adopted. The calculated energies were
adjusted again by observed wavelengths using the interactive
optimization procedure included in the ELCALC~\cite{RK69} program.
In the newest version of NIST database, about 48 experimental
energy levels belonging to the $2s2p, 2p^2, 2s3l, 2p3l, 2s4l,
2p4p, 2p4d, 2s5d$ and $2p5d$ configurations are listed. Here, we
reported energies of 350 levels belonging to the 28 configurations
of Si XI, namely $2s^2$, $2s2p$, $2p^2$, $2l3l'$, $2l4l'$,
$2l5l'$, $2l6l'$ and $2l7l'$~($l=s, p; l'= 0, 1, ..., n-1)$.

In Table 3 we list the level energies along with available
experimental and theoretical compilations for a comparison.
Present predictions show a good agreement with experimental ones
for a majority of levels. Even for the lowest-lying levels of
ground and lower excited configurations, the largest difference
does not exceed 2\%. Fig.1-(c) obviously indicates present results
being better than the theoretical data included in the Chianti.
For higher excited levels, we use the data of Coutinho \&
Trigueiros~\cite{CT01} to assess the accuracy, because Coutinho \&
Trigueiros~\cite{CT01} adjusted their results with observed
wavelengths. A good agreement (within 0.3\%) appears as shown by
the filled circled in Fig.3-(c) and in Table 3.

\section{3 Weighted oscillator strength}
We further reported the weighted absorption oscillator strength
($gf_{ij}$) and radiative decay rate ($A_{ij}$) for a given
transition $i\to j$ using the FAC package. In the following, we
discuss our results of weighted oscillator strengths for each of
the silicon ions in detail.

\subsection{3.1 Si IX}
Available theoretical data of radiative rates or weighted
oscillator strengths was confined to transitions among 46
low-lying levels~\cite{Agg98,BD93} of Si~IX. For strong or allowed
transitions, the oscillator strengths are accurate to 10\% as
stated by Aggarwal~\cite{Agg98} who adopted the CIV3
code~\cite{Cow81}. However larger uncertainties still exist for
those weak forbidden transitions ($gf<0.1$). In the work of
Orloski et al.~\cite{OTC99}, weighted oscillator strengths ($gf$)
to higher excited levels of $2s2p^23l$ configurations were also
reported. In addition, a majority of literatures reported the
radiative rates for electric dipole transitions among levels with
$n=2$ complexes. To our best knowledge, no additional works
reported the radiative rates or oscillator strengths for $\Delta
n\geq1$ transitions among levels of $n=2, 3, 4$ and 5 complexes.
In order to successfully modelling the astrophysical high-quality
spectra, here we extend the data of the radiative rates. The $gf$
and radiative rates for transitions among 560 levels were
calculated as listed in Table 10. Not only the E1 type transitions
are reported, but also other type transitions such as M1, E2 and
M2, have been calculated, which directly results in the number of
transitions increases by orders of magnitude.

In Table 4 we list some $gf$-values along with other available
data of Aggarwal~\cite{Agg98}, Orloski et al.~\cite{OTC99} and the
data included in the Chianti database for a comparison. The
Chianti code adopts results of Bhatia et al.~\cite{BD93} in the
modelling of astrophysical spectra. A visual comparison is shown
by Fig.2 for all available transitions. Because large CI has been
considered by Aggarwal~\cite{Agg98} and us, the two different
calculations show a good agreement (within 20\%) for most strong
transitions ($gf>0.1$) as shown by square symbols in Fig.2-(a).
Only three transitions such as $2s^22p^2~^1D_2$--$2s^22p3d~^3F_2$
(4--38, numbers corresponding to level indices),
$2s^22p3p~^1P_1$--$2s^22p3d~^1D_2$ (25--40) and
$2s^22p3p~^3D_1$--$2s^22p3d~^3F_2$ (26--38), show differences of
$>$20\%. Such differences are mainly due to the CI effect from
another three configurations of $3s3p$, $3p3d$ and $3d^2$
considered by Aggarwal~\cite{Agg98}. We gradually considered the
three configurations, the $gf$ increases from 0.2293 to 0.3031 for
$2s^22p^2~^1D_2$--$2s^22p3d~^3F_2$ (4--38) transition, whereas it
drops from 0.1995 and 0.1742 to 0.1454 and 0.1703 for
$2s^22p3p~^1P_1$--$2s^22p3d~^1D_2$ (25--40) and
$2s^22p3p~^3D_1$--$2s^22p3d~^3F_2$ (26--38) transitions,
respectively. This reveals that the CI is the most possible reason
for large uncertainties. The comparison with the data used by the
Chianti code is displayed by open-circle symbols in Fig.2-(a).
Though a less CI was considered in the work of Bhatia et
al.~\cite{BD93}, the two different data still agree within 20\%
for most strong transitions. Yet, for more transitions besides
above three ones, the two different calculations differ beyond
20\%. Additionally, the comparison with the results of Orloski et
al.~\cite{OTC99} is illustrated by up-triangle symbols in
Fig.2-(a). For those weak transitions ($gf<0.1$), more transitions
show large discrepancies between different calculations as shown
in Fig.2-(b). We also test the reason for the large differences,
and found that the CI effect is the main reason. Generally,
present results agree with the results of Aggarwal~\cite{Agg98},
whereas a slightly poor agreement with the data included in the
Chianti appears. The presence of large uncertainties suggests a
scope of improvement.
\begin{figure}
\includegraphics[angle=-90,width=7.5cm]{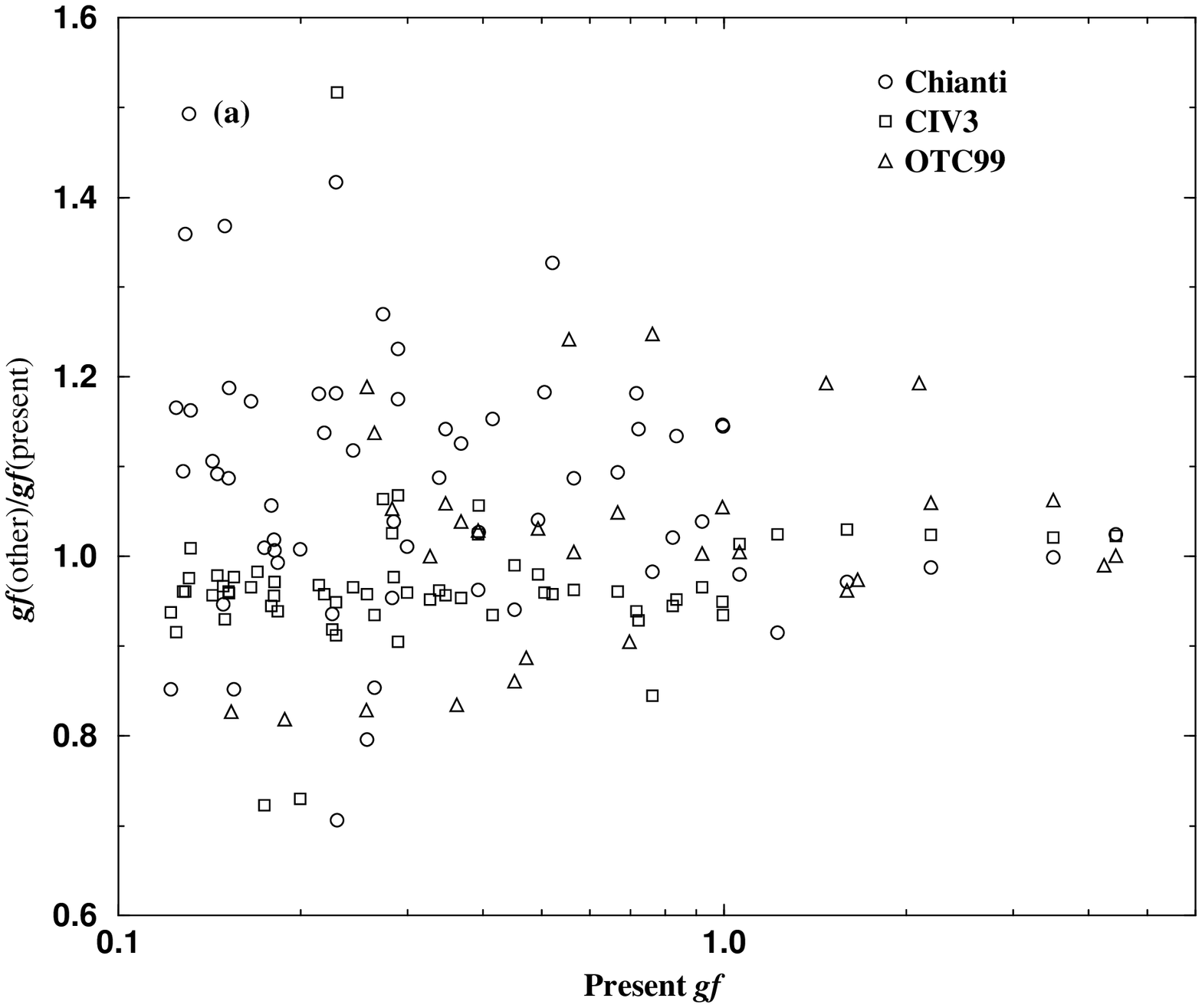}
\includegraphics[angle=-90,width=7.5cm]{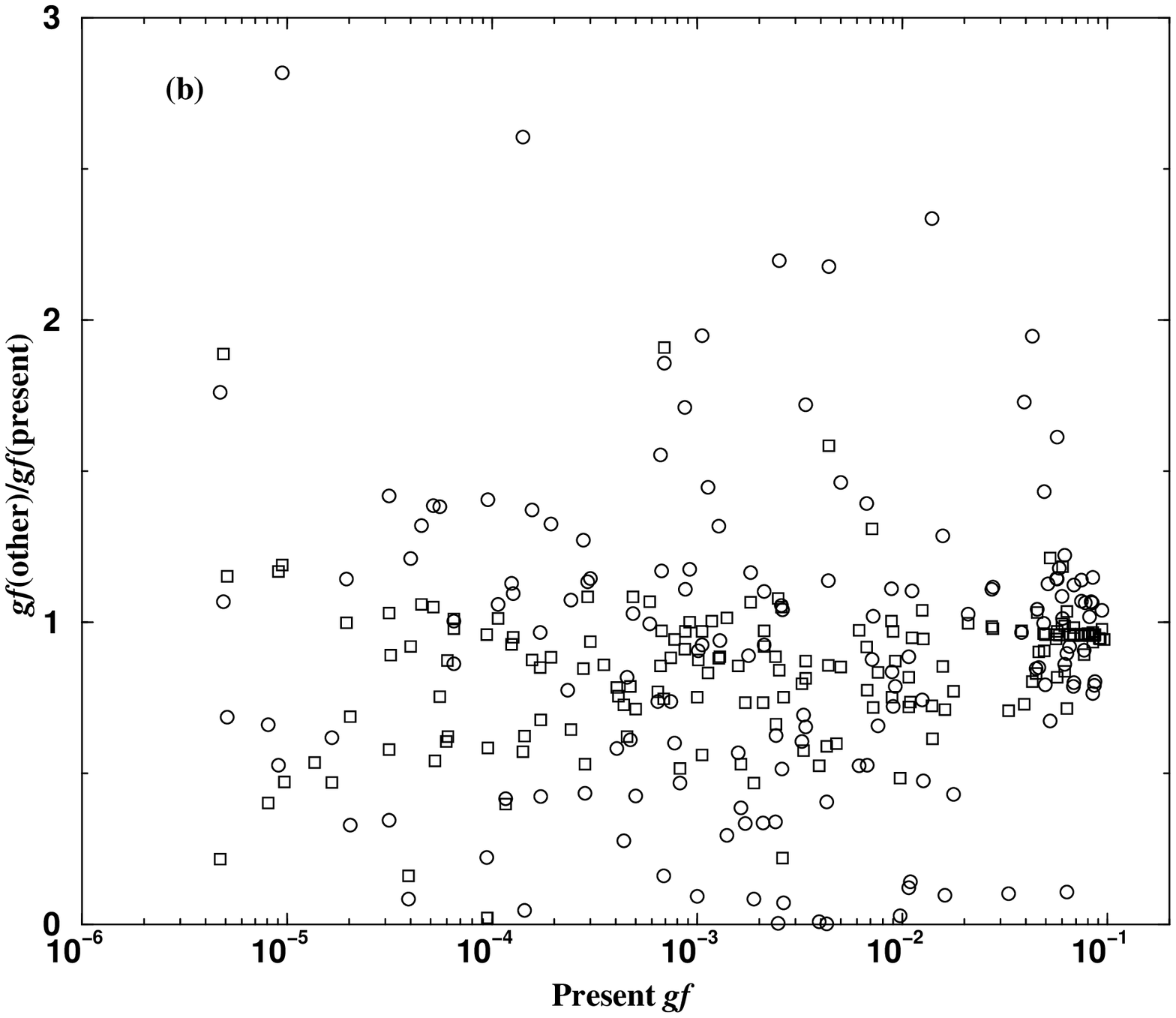}
\caption{Comparison of the Si~IX weighted oscillator strengths
from different calculations. The x-axis denotes the present
results, the y-axis denotes ratio of other theoretical predictions
{\it versus} present ones. (a) For strong transitions $gf>0.1$,
(b) For weak transition $10^{-6}<gf<0.1$.}
\end{figure}

\subsection{3.2 Si X}
For B-like silicon, most available radiative rates or oscillator
strengths of Si X are for transitions among the lowest 15 levels.
The radiative rates from higher excited levels were firstly
reported by Zhang $\&$ Sampson~\cite{ZS94}, which were extensively
used in current astrophysical modelling codes, such as Chianti and
APEC. Cavalcanti et al.~\cite{CLT00} calculated the weighted
oscillator strengths ($gf$) using optimized electrostatic
parameters for transitions among levels of the configurations,
namely $2s^22p$, $2p^3$, $2s^23l$, $2s2p3l$, $2p^23l$, $2s^24l$,
$2s2p4p$, $2s2p4d$, $2s^25d$, $2s2p5p$ and $2s2p5d$. In our study,
we calculated the $gf$ and radiative rates for transitions among
320 levels. By including other type transitions such as M1, E2 and
M2, the number of transitions increases by orders of magnitude.
These lost features in the astrophysical modelling codes might
explain the disagreement of fitting in astrophysical X-ray
spectral analyses. The results of $gf$ and radiative rates $A$ are
listed in Table 11, along with effective collision strengths which
will be discussed in the next section.

In Table 5 we compared our $gf$-values with other different
predictions for some transitions. Fig.3 visually exhibits such
comparison for all available transitions. Here we pay special
attention on the comparison with the data used by the Chianti
code. For most strong transitions, our results agree with
predictions of Zhang $\&$ Sampson~\cite{ZS94} within 20\%.
Fig.3-(a) clearly reveals that our results are systematically
higher than predictions used by the Chianti. However, three strong
transitions of $2s2p^2~^4P_{3/2}$--$2s2p(^3P)3d~^2D_{5/2}$
(4--48), $2s2p^2~^2D_{3/2}$--$2s2p(^3P)3d~^4D_{5/2}$ (7--45) and
$2p^3~^2P_{3/2}$--$2p^2(^3P)3d~^4D_{3/2}$ (15--44), display
discrepancies being up to a factor of $\sim$4. By gradually
decreasing the configurations, we confirm that the large
differences are mainly resulted from the CI effect of the $2p^3$
configuration. When only configurations of $2s^22p$, $2s2p^2$,
$2s^23l$ and $2s2p3l$ are considered, the $gf$ increase from 0.085
and 0.054 to 0.404 and 0.126 for the
$2s2p^2~^4P_{3/2}$--$2s2p(^3P)3d~^2D_{5/2}$ (4--48) and
$2s2p^2~^2D_{3/2}$--$2s2p(^3P)3d~^4D_{5/2}$ (7--45) transitions,
respectively, they show a good agreement with the values of the
Chianti. But the $gf$ rapidly decrease to 0.117 and 0.052 again,
when the $2p^3$ configuration has been included. For the
$2p^3~^2P_{3/2}$--$2p^2(^3P)3d~^4D_{3/2}$ (15--44) transition, the
large difference is due to the CI effect among the $n=3$
complexes. When only the $2p^23d$ configuration of the $n=3$
complexes has been included, the $gf$ is 0.1136 which agrees well
with the adopted data (0.1099) by the Chianti. When another two
related $3d$-complexes such as $3s^23d$ and $2s2p3d$ have been
included, the value increases to 0.1225. The $gf$ value steadily
increases to 0.1978 when other $n=3$ complexes have been included.
Therefore, the large differences between present results and other
different theoretical predictions, are due to the CI effect. For
those weak transitions ($gf<0.1$), more transitions show large
discrepancies between different calculations as shown in
Fig.3-(b). The differences are even up to an order of magnitude
for a few transitions, which leaves a scope of improvement.
\begin{figure}
\includegraphics[angle=-90,width=7.5cm]{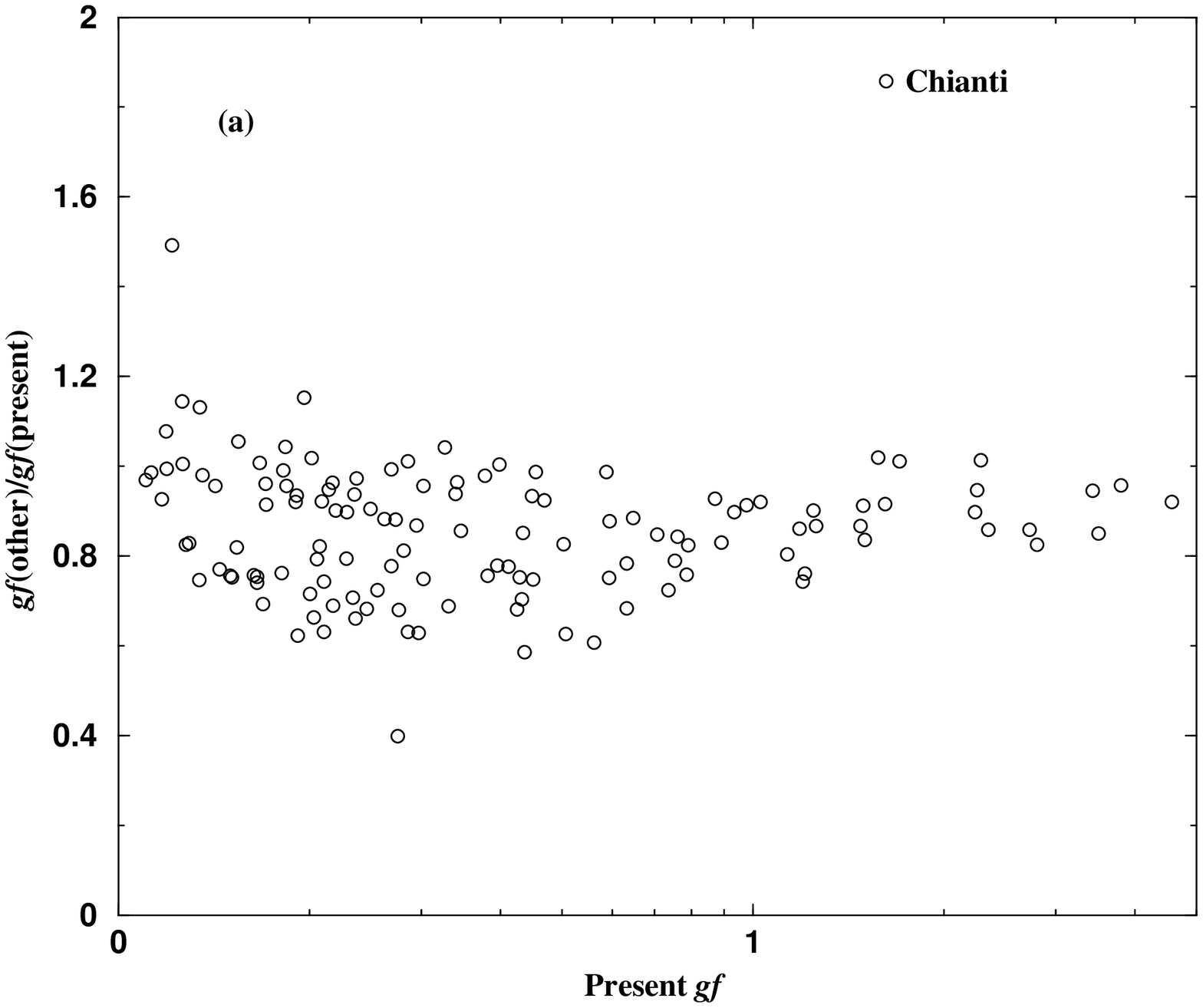}
\includegraphics[angle=-90,width=7.5cm]{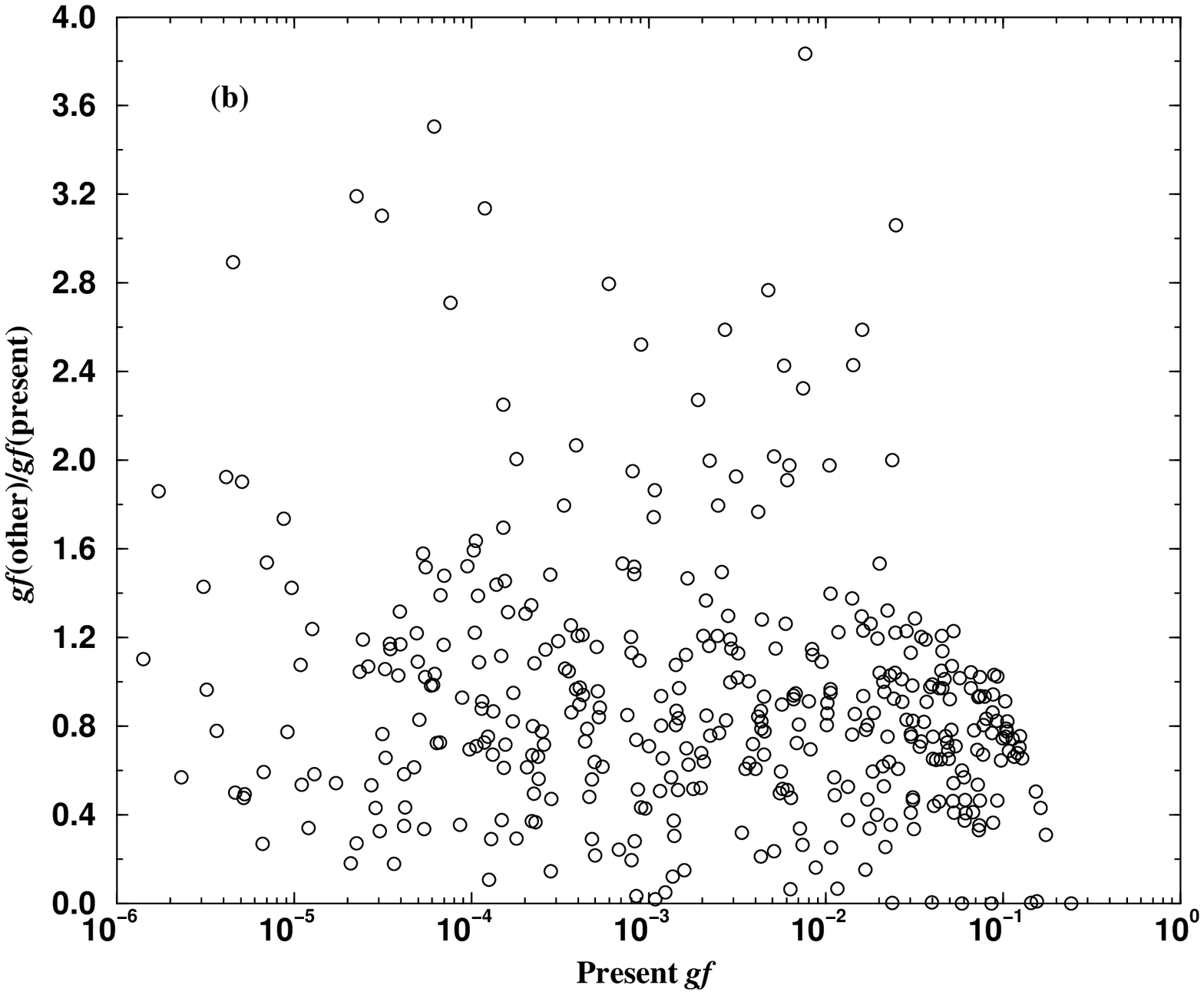}
\caption{Comparison of the Si~X weighted oscillator strengths from
different calculations. The x-axis denotes the present results,
the y-axis denotes ratio of the data included by the Chianti {\it
versus} present ones. (a) For strong transitions $gf>0.1$, (b) For
weak transition $10^{-6}<gf<0.1$.}
\end{figure}

\subsection{3.3 Si XI}
In the Chianti database, the $gf$ of electric dipole allowed
transitions of Si~XI are included for transitions among 46 levels
belonging to the $2s^2$, $2s2p$, $2p^2$, $2s3l$ and $2p3l$
configurations. These data was from study of Sampson et
al.~\cite{SG84} who adopted the DFS code developed by
them~\cite{SZM89}. Coutinho et al.~\cite{CT01} also performed the
calculation of the $gf$ by including another configurations of
$2s4l$, $2p4l$, $2s5l$ and $2p5l$. In our study, the $gf$ and
radiative rates for transitions among 350 levels were calculated.
By including other type transitions such as M1, E2 and M2, the
number of transitions increases by orders of magnitude relative to
available data. The radiative decay rates to higher excited levels
such as levels of $2l3l'$ configurations, have also been obtained.
These data could be used to investigate cascade effects to higher
levels, which may be important in the high-density laser-produced
plasma. The results of $gf$ and radiative rates $A$ are listed in
Table 12.

In Table 6 we compare our results with the data used by the
Chianti code and the results of Coutinho \& Trigueiros~\cite{CT01}
for some transitions. A visual comparison for all available
transitions is illustrated by Fig.4. For strong transitions
($gf>0.1$), present predictions agree with the data used by
Chianti code within $\sim$20\%. But three transitions such as
$2s^2~^1S_0$--$2s3p~^3P_1$ (1--15), $2s2p~^1P_1$--$2p3p~^1S_0$
(5--25) and $2s2p~^1P_1$--$2p3p~^3P_0$ (5--44), differ up to
$\sim$70\%. By gradually decreasing the configurations, we find
that the large difference is still present for the
$2s^2~^1S_0$--$2s3p~^3P_1$ (1--15) transition, and our prediction
is always significantly lower than the value used by the Chianti
when the $2p3l$ complexes are included. However, the $gf$ becomes
being higher than the later case by a factor of $\sim5$ without
the $2p3l$ complexes. In the work of Sampson et al.~\cite{SG84},
they explain that the large difference of collision strength
$\Omega$ to be a level mixing of $^1P_1$ and $^3P_1$. Whether the
same behavior occurs for the weighted oscillator strength $gf$. We
found that the sum $gf$ of the two transitions
$2s^2~^1S_0$--$2s3p~^1P_1$ (1--13) and $2s^2~^1S_0$--$2s3p~^3P_1$
(1--15), are 0.6068 and 0.6130 for above two cases, and this agree
with the sum $gf$ (0.5642) of Chianti within 10\%. In this study,
the sum $gf$ is 0.5892, which also agrees with that of Chianti.
Therefore the level mixing and CI maybe the possible reasons of
the large differences. For those weak transitions ($gf<0.1$), more
transitions show large discrepancies between different
calculations as shown in Fig.4-(b). The large differences may be
from the CI effect without considered, which leaves a scope of
improvement.
\begin{figure}
\includegraphics[angle=-90,width=7.5cm]{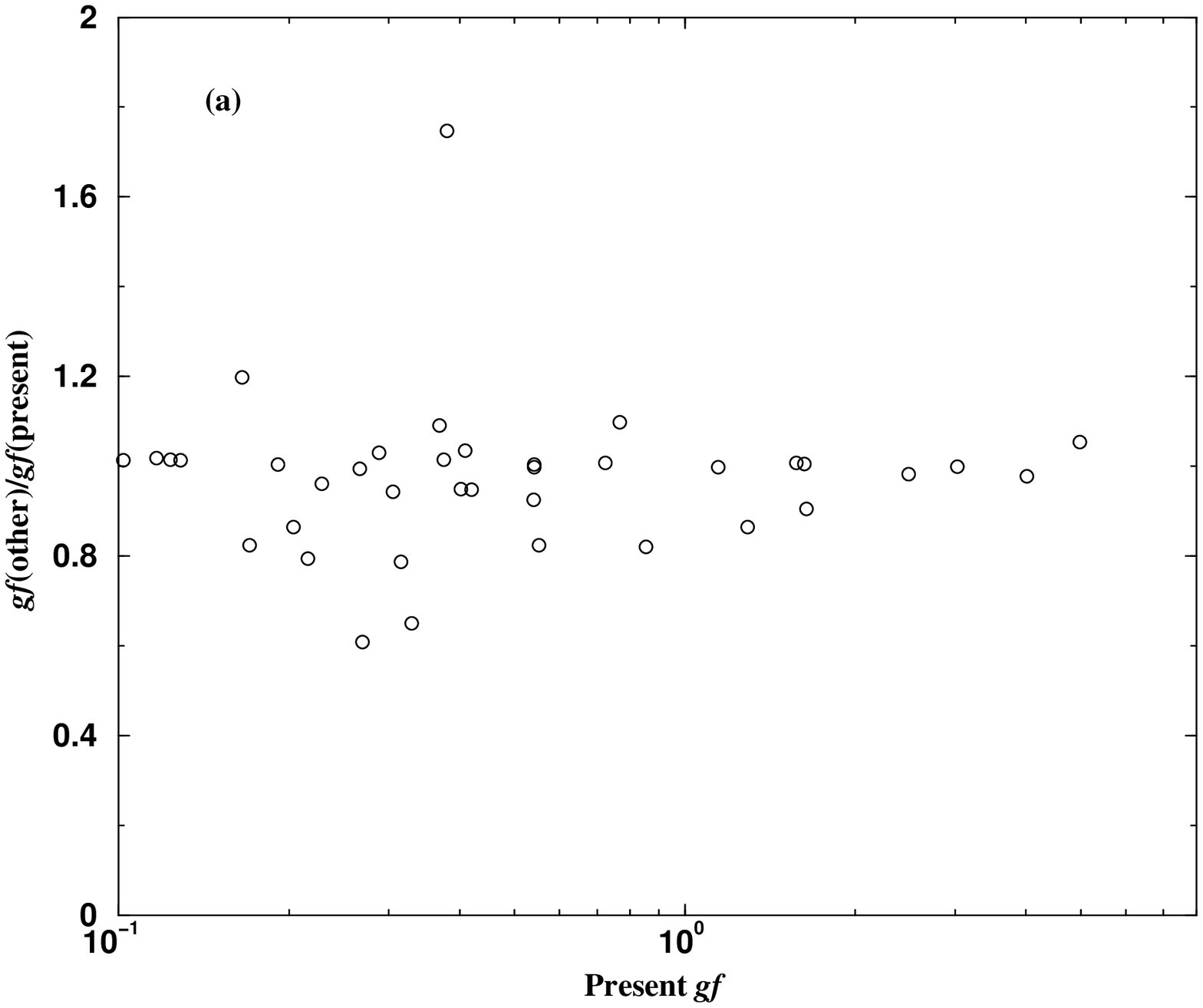}
\includegraphics[angle=-90,width=7.5cm]{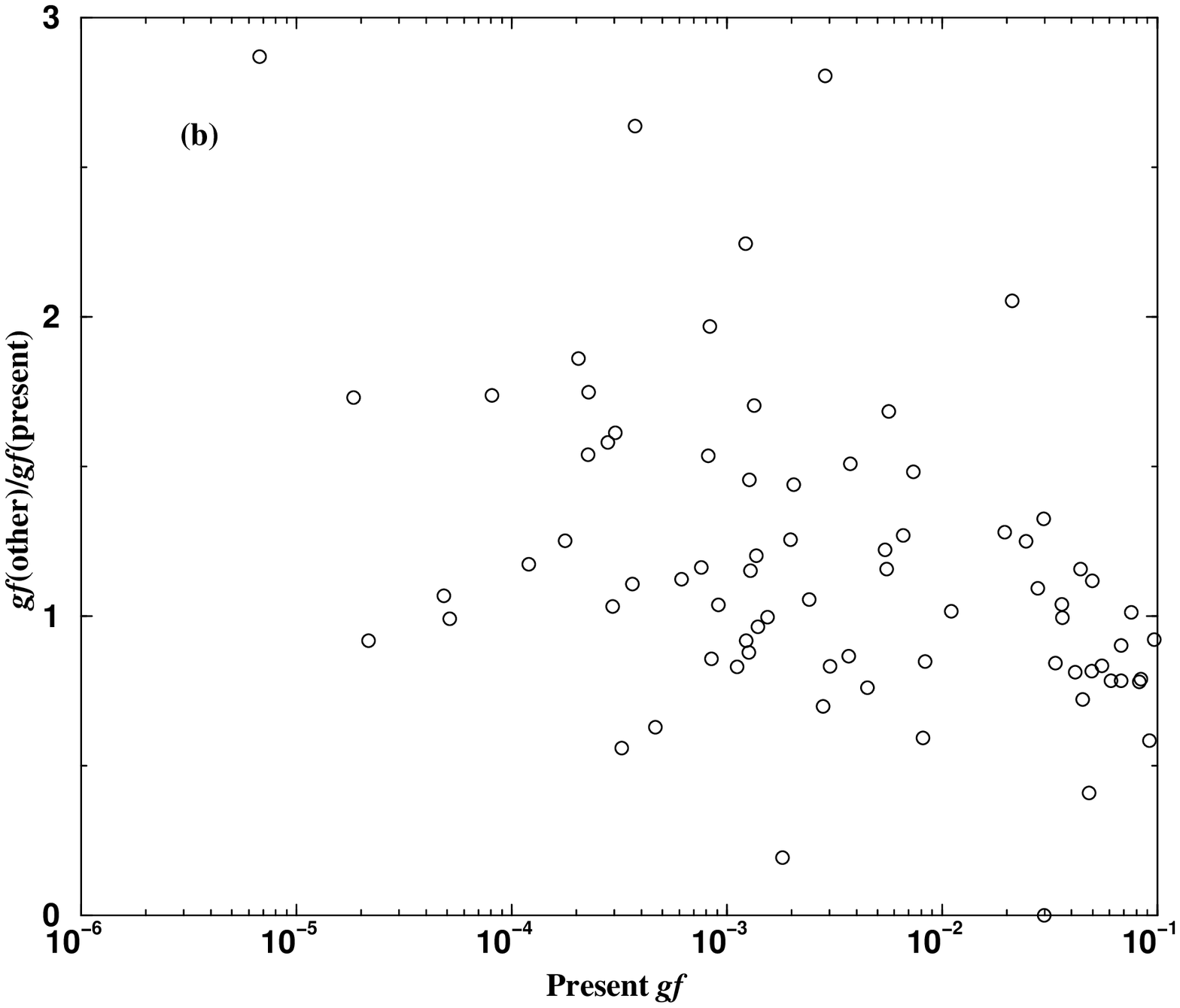}
\caption{Comparison of the Si~XI weighted oscillator strengths
from different calculations. The x-axis denotes the present
results, the y-axis represents ratio of the data included by the
Chianti {\it versus} present ones. (a) For strong transitions
$gf>0.1$, (b) For weak transition $10^{-6}<gf<0.1$.}
\end{figure}

\section{4 Collision strength $\Omega$ and effective collision strength $\Upsilon$}
In this study, a self-consistent electron impact collision
strengths $\Omega$ has been calculated at scattered electron
energies ranging from 4 to 10000~eV being cover the entire energy
range. In Table 7--9, we list the $\Omega$ at ten scattered
electron energy points such as 10, 50, 100, 200, 400, 600, 800,
1000, 1500 and 2000~eV. In practical applications, effective
collision strengths appear more important. Assuming a Maxwellian
distribution for the thermal electrons, the effective collision
strengths are derived through numerical integration,
\begin{eqnarray}
\Upsilon_{ij} & = & \int_0^{\infty}\Omega_{ij}{\rm
exp}\left(-\frac{E}{kT_e}\right)d\left(\frac{E}{kT_e}\right),
\nonumber
\end{eqnarray}
where $E$ is the scattered electron energy, $k$ is the Boltzmann
constant, and $T_e$ is the electron temperature. Correspondingly,
the excitation rate coefficients (in cm$^3$s$^{-1}$) used in the
statistical equilibrium to derive the level populations, can be
obtained directly from the effective collision strength
\begin{eqnarray}
C_{ij} & = & \frac{8.63\times10^{-6}}{\omega_iT_e^{1/2}}{\rm
exp}\left(\frac{-E_{ij}}{kT_e}\right)\Upsilon_{ij}, \nonumber
\end{eqnarray}
where $E_{ij}$ is the energy difference between levels $i$ and
$j$, and $\omega_i$ is the statistical weight of level $i$. In the
following, we present our results of (effective) collision
strengths for each ions in detail, and make an assessment through
comparison with published data.

\subsection{4.1 Si IX}
In the Chianti database, only limited excitation data among the
lowest 46 levels is available for Si IX, and the data are taken
from work of Bhatia \& Doschek~\cite{BD93} who reported values at
three incident electron energies of 20, 40 and 60~{\it Ryd}. The
work of Aggarwal \& Baluja~\cite{Agg83} for forbidden transitions
is also included. Aggarwal \& Baluja~\cite{Agg83} adopted a more
accurate approach---$R-$matrix~\cite{Hib75} to derive the
excitation data, so far, which is the best reliable data for this
ion, because coupling effects among different channels have been
considered. But only the excitations among levels with $n=2$
complexes were reported. A much earlier work is performance of
Mason \& Bhatia~\cite{MB78} who calculated the excitation data
among the lowest 20 levels using the distorted wave approximation
(DWA). To our knowledge, the latest work of the collision
strengths for carbon-like Si is results of Zhang \&
Sampson~\cite{ZS96}, yet only $\Delta n=0$ DWA excitation data
within $n=2$ configurations is available. Here we present a
self-consistent excitation data for a large amount of transitions.

In Table 10, we present the effective collision strengths for
excitations from lowest 20 levels to higher levels up to 560-th
level at seven temperatures: 0.5, 1.0, 2.0, 3.0, 4.0, 5.0 and
6.0~MK. Moreover the weighted oscillator strengths ($gf$) and
radiative rates ($A$) are listed in this table. The indices used
to represent the lower and upper levels of a transition have
already been defined in Table 1.

\begin{figure}
\includegraphics[angle=-90,width=7.6cm]{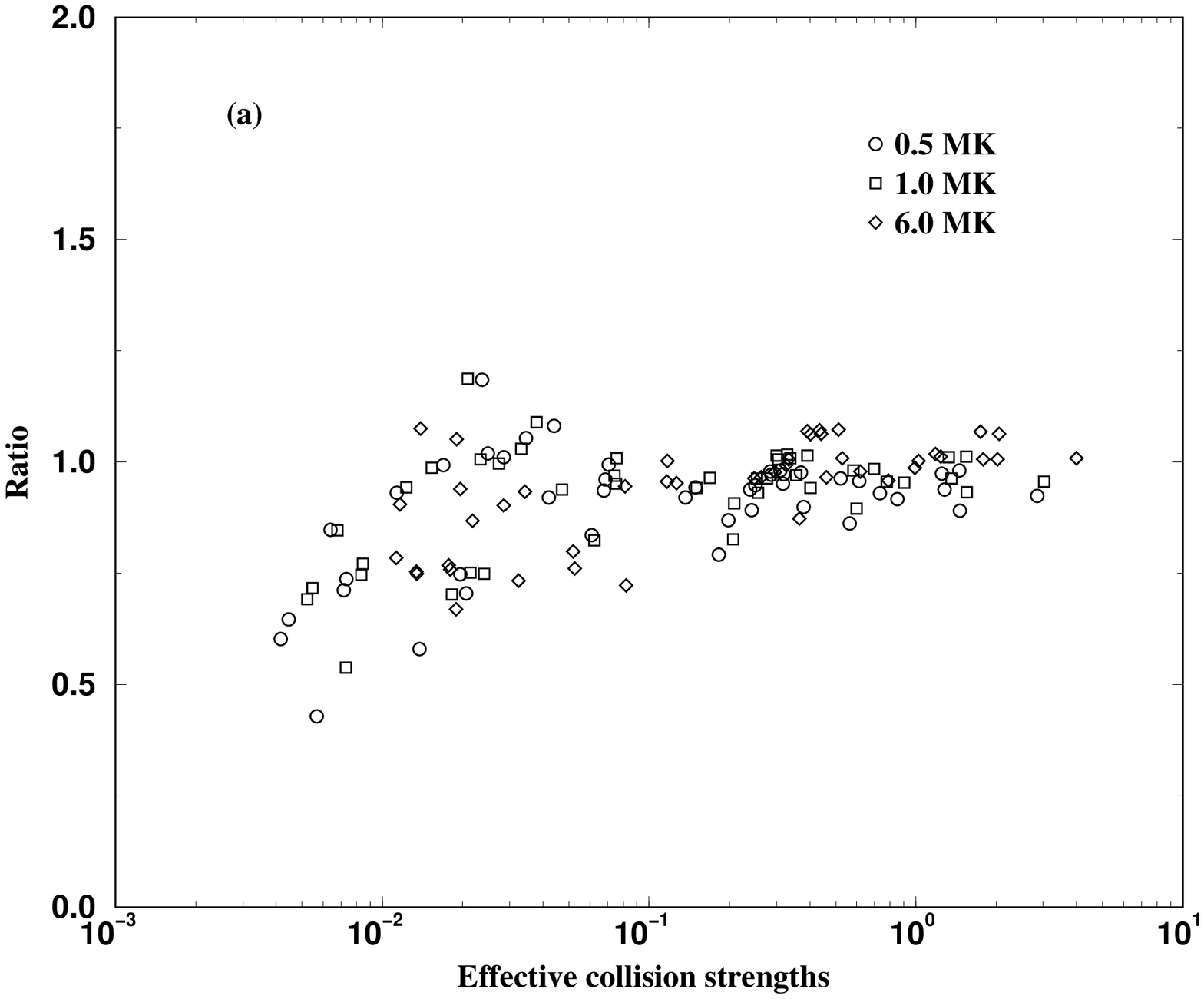}
\includegraphics[angle=-90,width=7.5cm]{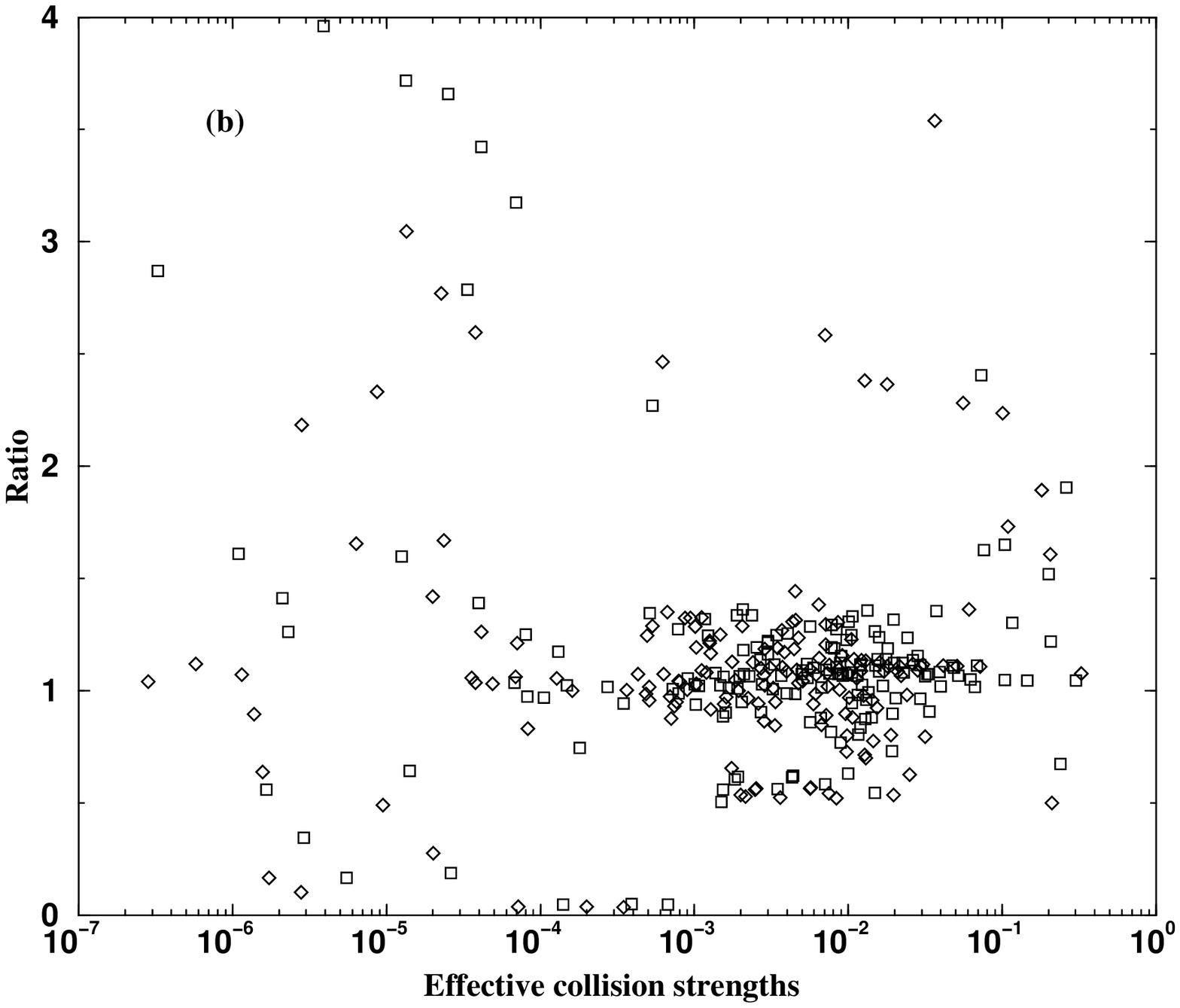}
\caption{Comparison of the effective collision
strengths~$\Upsilon$ of Si~IX for different calculations. The
x-axis denotes the present $\Upsilon$, the y-axis denotes ratio of
the data used by the Chianti code {\it versus} present ones. (a)
allowed transitions, while (b) other type transitions.}
\end{figure}
We pay a special attention on the comparison with the data used by
the Chianti code. In Table 13, the $\Upsilon$ is given for some
transitions at three temperatures of 0.5, 1.0 and 6.0~MK. The
comparison with all available $\Upsilon$ is shown by Fig.5.
Fig.5-(a) illustrates the comparison for allowed transitions,
while Fig.5-(b) displays the comparison for other type
transitions. For the allowed transitions, present results of
$\Upsilon$ are systematically higher than the data used by the
Chianti code slightly, yet they agree within $\sim$20\% except for
a few excitations to the levels of $2s^22p3s$ such as
$2s^22p^2~^3P_0$--$2s^22p3s~^3P_1$ (1--22),
$2s^22p^2~^3P_1$--$2s^22p3s~^3P_0$ (2--21) and
$2s^22p^2~^1D_2$--$2s^22p3s~^1P_1$ (4--24), in which the
differences are up through to 50\%. For these transitions, the
$gf$ differs $\sim$30\% between the two different theoretical
calculations. Discussions in Sect.2.1 have indicated that the CI
maybe the possible reason. For forbidden transitions, we note that
the discrepancies are even higher up to factors of $\sim$1.5--9.
The $\Upsilon$ of the Chianti is systematically higher than
present results for excitations up through to the level of
$2p^4~^1S_0$. And the differences can reach up to $\sim$4 times in
some cases such as $2s^22p^2~^3P_0$--$2s^22p^2~^3P_1$ (1--2),
$2s^22p^2~^3P_0$--$2s^22p^2~^3P_2$ (1--3),
$2s^22p^2~^3P_1$--$2s^22p^2~^3P_2$ (2--3), etc. This is due to the
the consideration of resonant excitations in the calculation of
Aggarwal \& Baluja~\cite{Agg83} who adopted the $R-$matrix method.
In some forbidden transitions from much higher excited levels
($>20$), we also note the discrepancies being up to an order of
magnitude, such as $2s2p^3~^3D_3$--$2s^22p3p~^3S_1,~^3P_{1,2}$
(7--29, 31, 32). Such differences follow the discrepancies in the
relevant $gf$, in which the CI effect gives rise to the large
differences. However for most forbidden excitations, present
$\Upsilon$ is in agreement with the data used by the Chianti code.
This discussion indicates that there is a scope of improvement by
consideration of resonant effect and much more CI effect.

\subsection{4.2 Si X}
So far, only two literatures reported $\Omega$ and/or $\Upsilon$
of $n=2$--3 excitations, one is the work of Zhang $\&$
Sampson~\cite{ZS94} using relativistic DW approximation, the other
is an unpublished calculation of Sampson $\&$ Zhang~\cite{SZ95}.
The two calculations are extensively used by present astrophysical
modelling. Zhang et al.~\cite{ZGP94} had adopted $R$-matrix
approach to calculate the $\Omega$, whereas their calculation is
confined to transitions among 15 fine-structure levels belonging
to 8 LS terms $2s^22p(^2P^0_{1/2,3/2})$,
$2s2p^2$($^4P_{1/2,3/2,5/2}$, $^2D_{3/2,5/2}$, $^2S_{1/2}$,
$^2P_{1/2,3/2}$), $2p^3(^4S^0_{3/2}$, $^2D^0_{3/2,5/2}$,
$^2P^0_{1/2,3/2})$, and only partial data was reported. Recently,
Keenan et al.~\cite{KOT00} re-calculated these data again using
$R$--matrix method, and listed all these data at much finner
temperature grids. These excitation data was widely used by
current astrophysical modelling codes such as Chianti, MEKAL and
APEC codes. However a poor modelling for emissions of highly
charged Si in the astrophysical spectral analysis, is still
existed~\cite{GAB01,ABG01}. As stated in these literatures,
uncertainties of the excitation data is the main possible source
of the poor modelling. In our present work, a self-consistent
calculation is reported for the collision strength at ten
scattered electron energies as listed in Table 8. Assuming the
Maxwellian energy distribution, averaged collision strengths
($\Upsilon$) at seven temperatures are listed in Table 11 for
excitations from lowest 15 levels. The indices used to represent
the lower and upper levels for a given transition have already
been defined in Table 2.

\begin{figure}
\includegraphics[angle=-90,width=7.5cm]{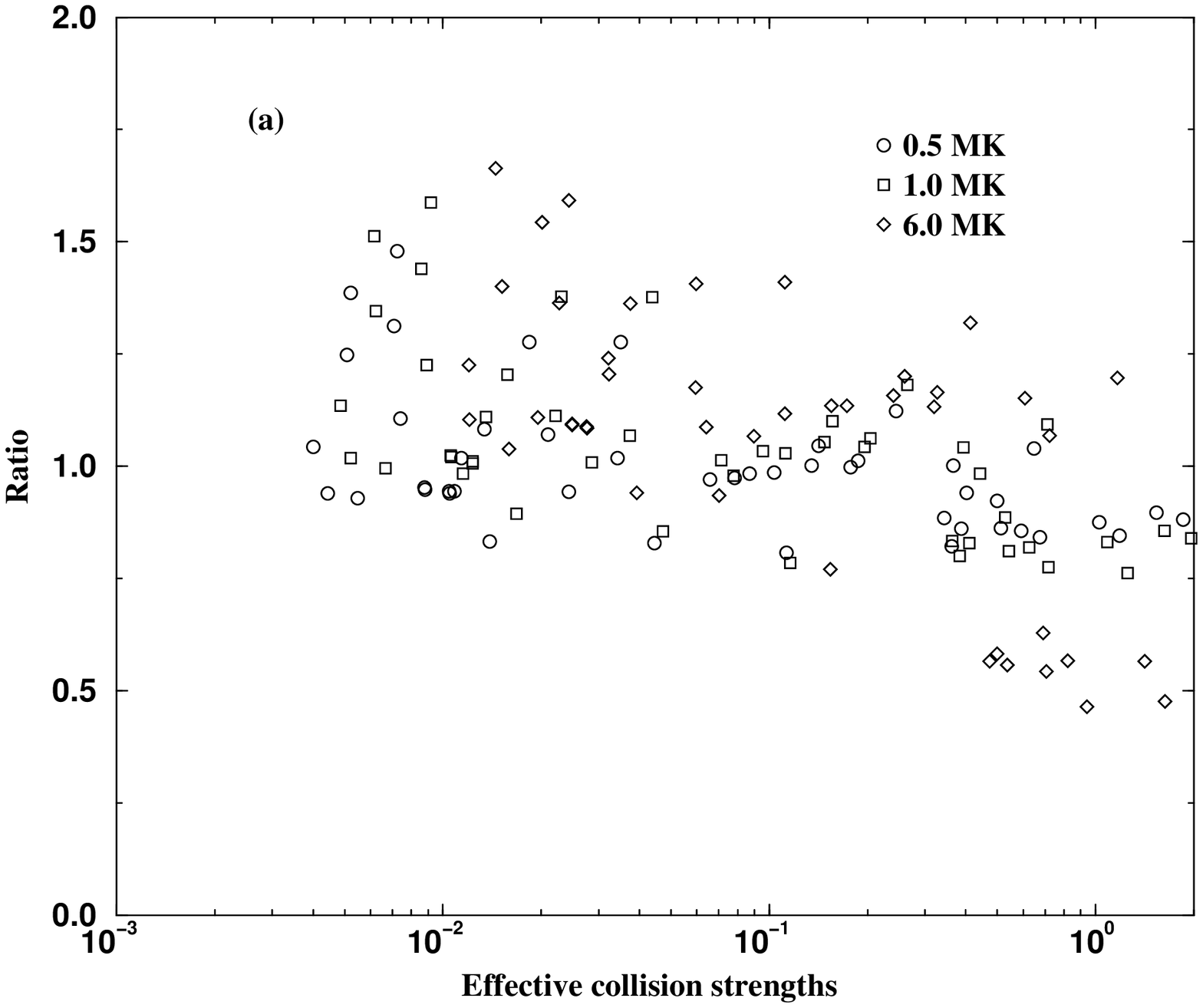}
\includegraphics[angle=-90,width=7.5cm]{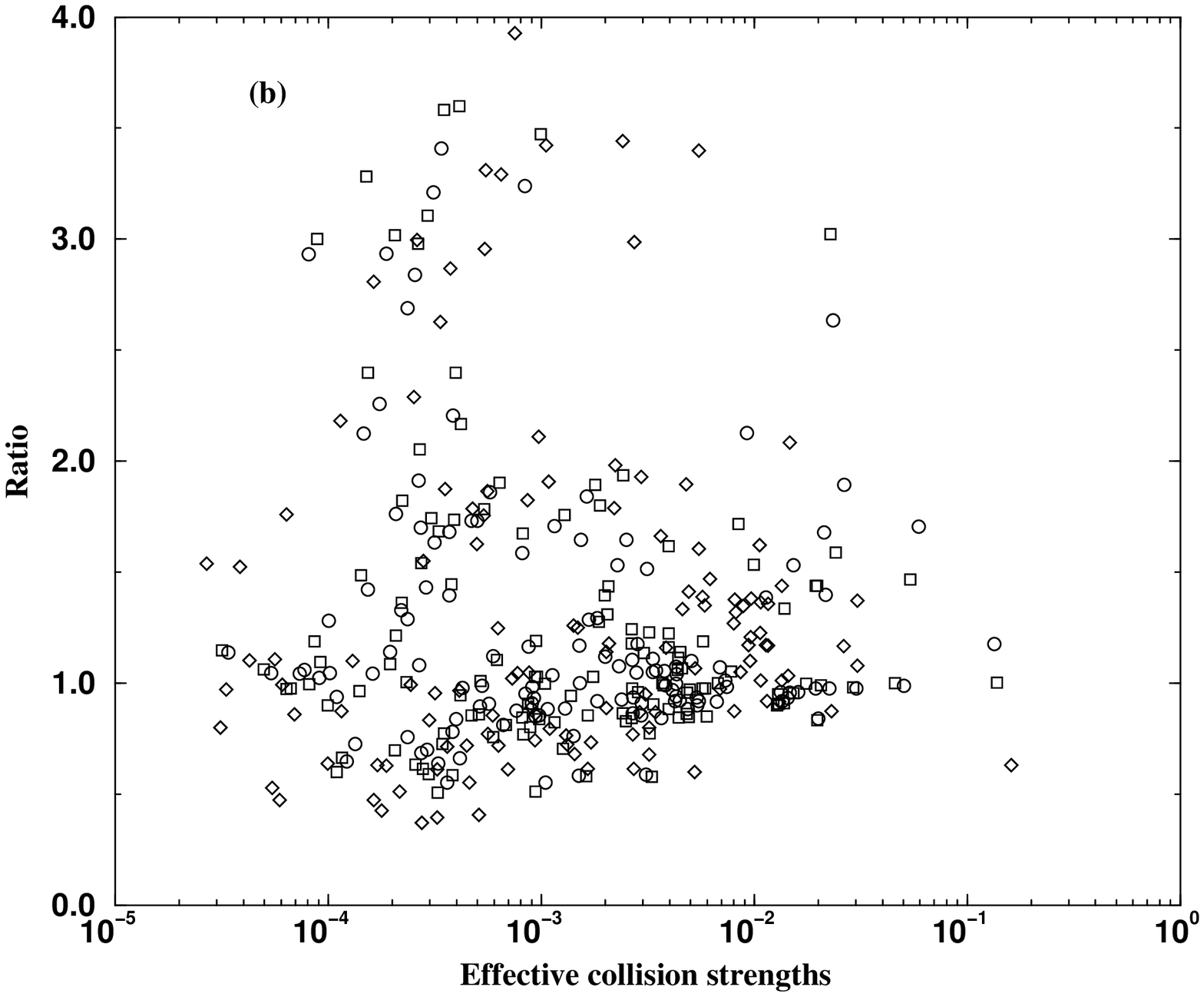}
\caption{Comparison of the effective collision
strengths~$\Upsilon$ of Si~X for different calculations. The
x-axis denotes the present $\Upsilon$, the y-axis denotes ratio of
the data used by the Chianti code {\it versus} present ones. (a)
allowed transitions, while (b) other type transitions.}
\end{figure}
In Table 14, we compare our $\Upsilon$ with the data used by the
Chianti code at three temperatures of 0.5, 1.0 and 6.0~MK, which
are typical temperatures of stellar X-ray emitters. The comparison
with all available $\Upsilon$ is illustrated in Fig.6. Fig.6-(a)
represents the allowed transitions, while Fig.6-(b) shows the
comparison for other type transitions. For most allowed
transitions, present results of $\Upsilon$ agree with the data
included in the Chianti data sets within 20\%, yet differences up
to 50\% are also apparent, such as excitations of
$2s^22p~^2P_{1/2}$--$2s2p^2~^2D_{3/2}$ (1--7),
$2s^22p~^2P_{3/2}$--$2s2p^2~^2D_{5/2}$ (2--6), etc. We found that
the large differences occur for excitations up through to the
level of $2s2p^3~^2P_{3/2}$. And at higher temperature, the
differences is more clear. In the Chianti database, the work of
Zhang et al. ~\cite{ZGP94} is used for these excitations, who
adopted $R-$matrix method by accounting for the resonant effect.
So we believe the large differences are resulted from the resonant
effect. For those forbidden transitions, more excitations exhibit
large differences. And in some cases, the discrepancies are up
through to factors of $\sim$4. Such phenomena is natural, because
the resonant excitations generally is more obvious. Fox example,
$2s^22p~^2P_{3/2}$--$2s2p^2~^4P_{1/2}$ (2--3), the result of Zhang
et al.~\cite{ZGP94} is higher than present one by a factor of
$\sim$2. While for an allowed transition of
$2s^22p~^2P_{1/2}$--$2s2p^2~^2S_{1/2}$ (1--8), the difference is
less than 20\%. In addition, the CI effect is another important
reason for the large differences, such as for
$2s2p^2~^4P_{3/2}$--$2s2p(^3P)3d~^2D_{5/2}$ (4--48) and
$2s2p^2~^4P_{3/2}$--$2p^2(^3P)3s~^2P_{1/2}$ (4--70). However, such
differences follow the discrepancies of the relevant $gf$. In the
Sect.2.2, we distinguish the large differences are from the CI
effect of the configuration $2p^3$. This comparison also indicates
that the improvement by considering the resonance excitation and
more CI effect, is very necessary.

\subsection{4.3 Si XI}
For this ion, only limited electron impact excitation data among
the lowest 46 levels is available so far, and all excitation data
was calculated using DW approximation. Yet most available $\Omega$
is confined to transitions with $\Delta n=0$. To our best
knowledge, only one work performed the calculation for $\Delta
n=1$ transitions, and it is a work about two decades
ago~\cite{SG84}. These data is still used by the Chianti code. In
this study, self-consistent results of $\Omega$ and $\Upsilon$ of
excitations from lowest 10 levels, are reported for Si XI as shown
in Table 9 and 12 respectively. The reported $\Upsilon$ covers the
typical coronal temperature range of $\sim$0.1--10~MK. The indices
used in tables to represent the lower and upper levels of a
transition have already been defined in Table 3.

\begin{figure}
\includegraphics[angle=-90,width=7.5cm]{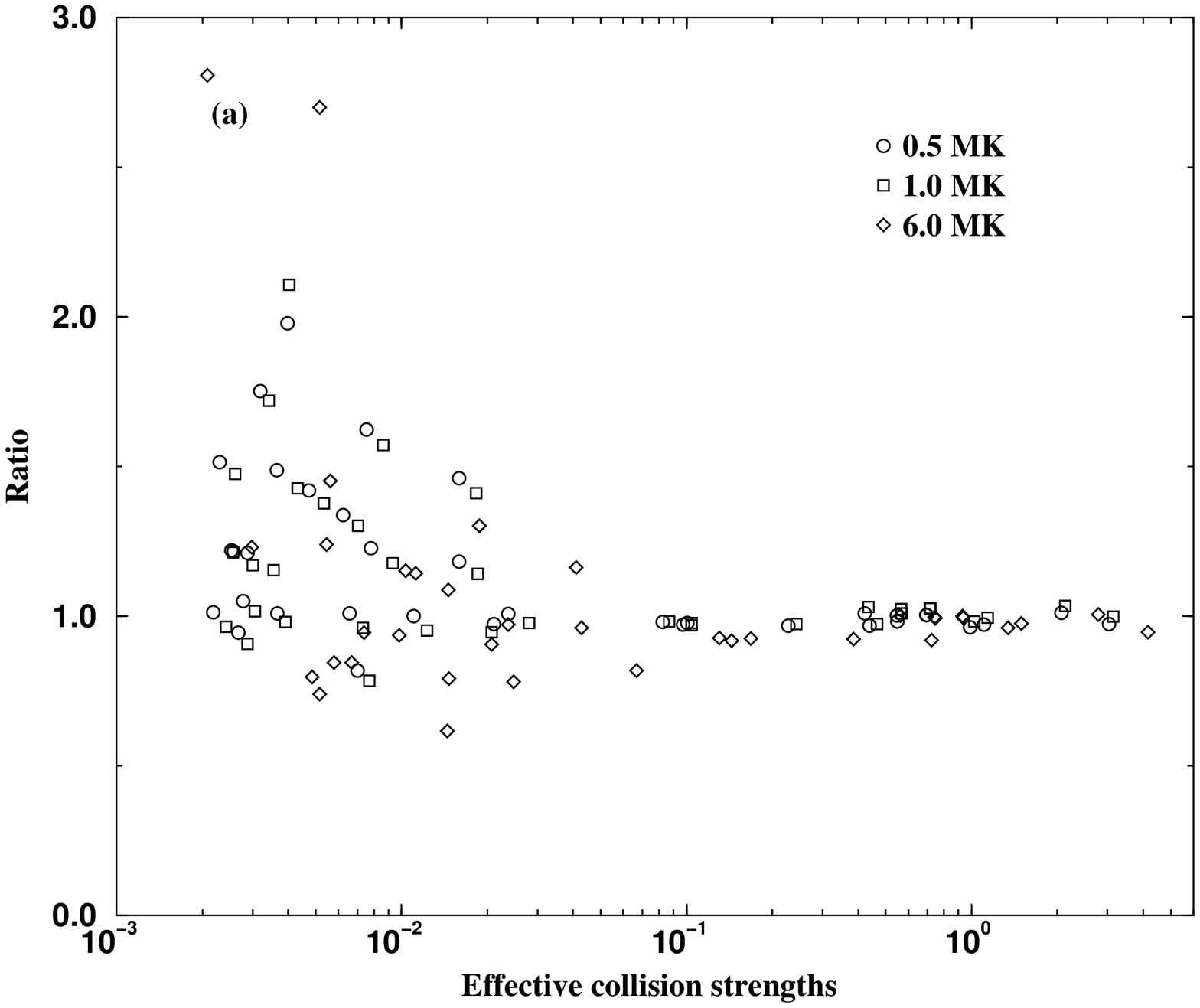}
\includegraphics[angle=-90,width=7.5cm]{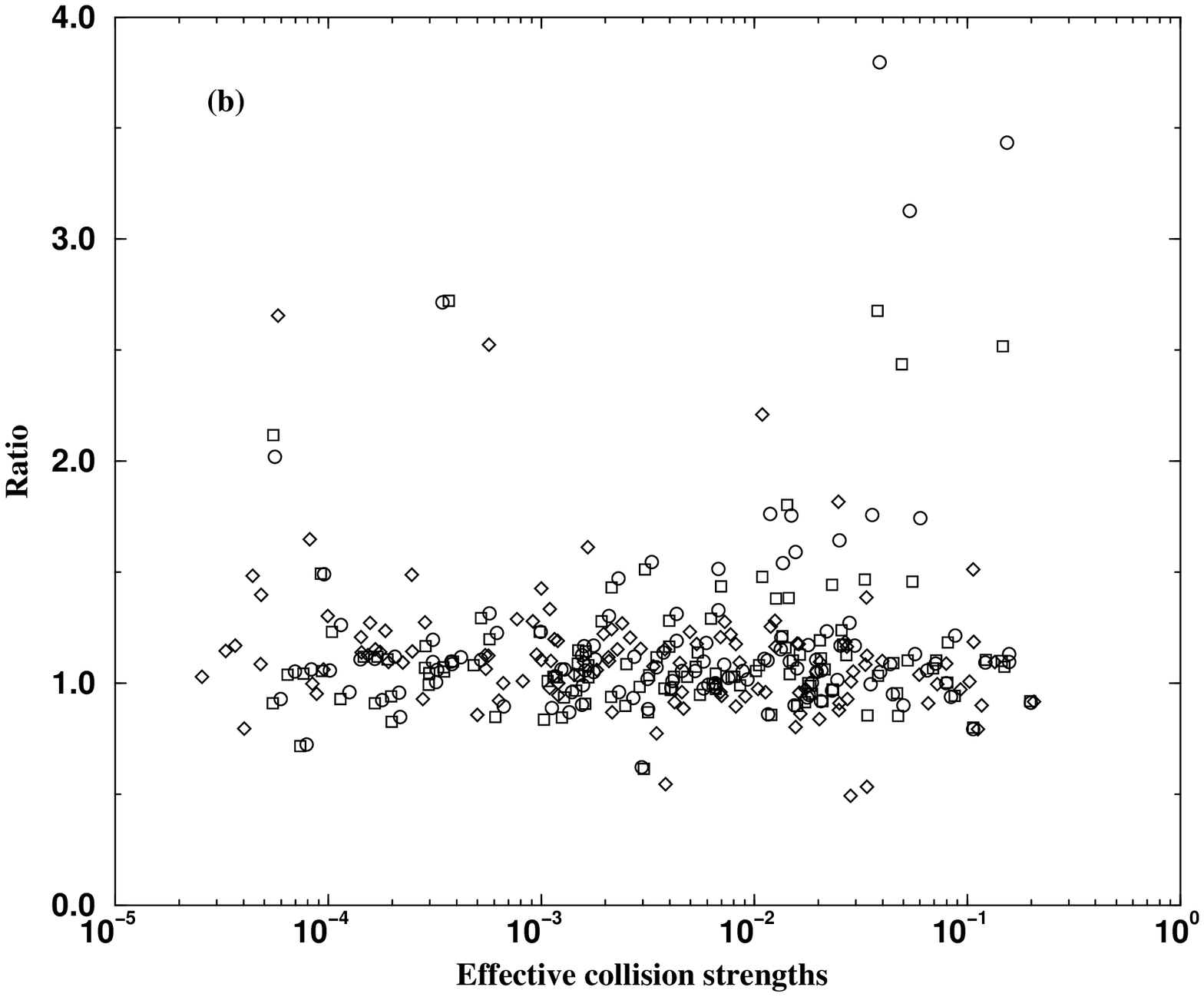}
\caption{Comparison of the effective collision
strengths~$\Upsilon$ of Si~XI for different calculations. The
x-axis denotes the present $\Upsilon$, the y-axis denotes ratio of
the data used by the Chianti code {\it versus} present ones. (a)
allowed transitions, while (b) other type transitions.}
\end{figure}
In Table 15, a comparison of our $\Upsilon$ with the data used by
the Chianti code at three temperatures is given for partial
transitions. The comparison with all available $\Upsilon$ is also
illustrated through Fig.7. Fig.7-(a) illustrates the comparison
for allowed transitions at three temperatures of 0.5, 1.0 and
6.0~MK, while Fig.7-(b) shows the comparison for other type
transitions. For the allowed transitions, the two different
calculations show a better agreement except for a few transitions
such as $2s^2~^1S_0$--$2p3s~^3P_1,~^1P_1$ (1--22, 24), and
$2s^2~^1S_0$--$2p3d~^3D_1$ (1--38). In these cases, the
differences are up to a factor of $\sim$4, and they follow the
differences in $gf$. The discussion in Sect.2.3 indicates the
discrepancies in $gf$ are due to the level mixing and CI effect.
For other type transitions, a majority of transitions shows a good
agreement at temperature of $\sim$0.1--6~MK, whereas large
differences are also present for some transitions. The
interpretation for the large discrepancies can be divided into two
group, one is due to the different data sources in the Chianti
data~sets. For excitations among the low-lying 10 levels, the work
of Berrington et al.~\cite{BBD85} is used, who considered the
resonant effect by using the $R$-matrix approach. So the
$\Upsilon$ of the Chianti is larger than present ones by factors
of up through to $\sim$3. The other possible reason maybe from the
different inclusions of CI, such as for $2s^2~^1S_0$--$2p3s~^3P_1$
(1--22). The difference also follows the discrepancy in relevant
$gf$. This leaves a scope of improvement by considerations of
resonance effect and much more CI effect.

\section{5 Conclusions}
In this work, radiative rates and oscillator strengths $gf$ among
560, 320 and 350 levels of Si IX, Si X and Si XI are reported,
respectively. A self-consistent calculation of (effective)
collision strengths is performed with large CI using the fully
relativistic FAC code of Gu~\cite{Gu03}. In general, our energy
levels agree with the available experimental values within 2\%.
For strong transitions ($gf>0.1$), present results of radiative
rates or weighted oscillator strengths show a good agreement with
available data except for a few transitions with discrepancies of
several times. By gradually decreasing the CI, we found that the
relatively large differences can be diminished, which indicates CI
plays an important role in accurate determinations of $gf$. For
weak transitions ($gf<0.1$), the CI effect appears more obvious.
And the discrepancies can reach up to an order of magnitude in
some cases.

Self-consistent collision strengths and effective collision
strengths by considering large CI are also reported in this study.
For allowed transitions, present work shows a better agreement
with available data except for a few excitations, in which the
differences are up to factor of $\sim$3. Such differences follow
the discrepancies of $gf$, which reveals the large differences
maybe from the CI effects. Though the present $\Upsilon$ agree
with the data of the Chianti within 20\% for a majority of
forbidden transitions, a certain of excitations show large
differences up to $\sim$4, which can be attributed to the resonant
effect and CI effect in the different calculations, because the
excitation data among levels of $n=2$ complexes in the Chianti
data~sets was from $R-$matrix method. The discussion also implies
that there is a scope of improvement by considering the resonant
effect and more CI effect.

\section{acknowledgments} This work was supported by the National
Natural Science Foundation under Grant No. 10433010, No. 10373014
and No. 10403007. This work is also financed by Chinese Academy of
Sciences under Grant No. KJCX2-W2.

\setcounter{table}{3}
\begin{table*}
\caption{Comparison of weighted oscillator strength $gf_{ji}$
(written in form a$\pm$b implying a$\times10^{\pm b}$) for some
transitions of Si IX. $i$ and $j$ are denoted in Table I. OTC99
represents the work of Orloski et al.~\cite{OTC99}.}
\begin{ruledtabular}
\begin{tabular}{rrrrrrrrrrrr}
$i$ & $j$ & FAC &CIV46 & Chianti & OTC99 & $i$ & $j$ & FAC & CIV46 & Chianti & OTC99 \\
            \hline
   1 &  9 & 8.160-2 & 7.797-2 & 8.310-2 & 7.797-2 &   3 & 38 & 1.251-2 & 1.301-2 & 9.310-3 &  1.301-2\\
   1 & 11 & 7.759-2 & 7.444-2 & 8.260-2 & 7.444-2 &   3 & 39 & 4.623-2 & 4.170-2 & 3.930-2 &  4.170-2\\
   1 & 14 & 9.369-2 & 9.154-2 & 9.750-2 & 9.154-2 &   3 & 40 & 5.010-3 & 4.274-3 & 7.330-3 &  4.274-3\\
   1 & 22 & 6.791-2 & 6.494-2 & 5.360-2 & 6.494-2 &   3 & 42 & 6.889-4 & 1.315-3 & 1.280-3 &  1.315-3\\
   1 & 24 & 6.702-4 & 6.514-4 & 7.840-4 & 6.514-4 &   3 & 43 & 7.678-2 & 6.860-2 & 6.970-2 &  6.860-2\\
   1 & 42 & 1.061+0 & 1.076+0 & 1.040+0 & 1.076+0 &   3 & 44 & 3.494+0 & 3.566+0 & 3.490+0 &  3.566+0\\
   1 & 46 & 6.041-2 & 7.154-2 & 6.120-2 & 7.154-2 &   3 & 46 & 3.925-1 & 4.024-1 & 3.780-1 &  4.024-1\\
\end{tabular}
\end{ruledtabular}
\end{table*}

\setcounter{table}{4}
\begin{table*}
\caption{Comparison of weight oscillator strength ($gf$) for some
transitions of Si X. CLT00 refers to the work of  Cavalcanti et
al.~\cite{CLT00}.}
\begin{ruledtabular}
\begin{tabular}{rrrrrrrrrrrr}
$i$ & $j$ & FAC &  CLT00 & Chianti & $i$ & $j$ & FAC &  CLT00 & Chianti \\
 \hline
  1 &  7 & 1.421-1 &         & 1.360-1 & 4 & 21 & 1.500-1 &         & 1.134-1\\
  1 &  8 & 1.126-1 &         & 1.110-1 & 4 & 45 & 2.807+0 & 2.380+0 & 2.316+0 \\
  1 &  9 & 1.900-1 & 1.840-1 & 1.750-1 & 4 & 53 & 8.366-4 &         & 2.357-4 \\
  1 & 16 & 5.113-2 &         & 4.013-2 & 4 & 56 & 5.504-4 &         & 3.396-4 \\
  1 & 19 & 1.246+0 & 1.300+0 & 1.124+0 & 4 & 74 & 1.172-3 &         & 9.430-4 \\
  1 & 28 & 1.200-1 & 1.870-1 & 8.156-2 & 4 & 82 & 5.016-2 &         & 4.633-2 \\
  1 & 59 & 5.940-4 &         & 1.661-3 & 5 & 25 & 5.520-5 &         & 8.375-5 \\
  1 &144 & 4.307-2 & 8.930-2 &         & 5 & 45 & 2.558-1 & 6.400-3 &  \\
  2 &  6 & 2.373-1 & 2.320-1 & 2.040-2 & 5 & 49 & 1.575+0 & 1.900+0 & 1.606+0 \\
  2 &  9 & 1.545-1 & 1.570-1 & 1.630-1 & 5 & 66 & 3.527-4 &         & 3.698-4 \\
  2 & 10 & 5.872-1 & 5.820-1 & 5.800-1 & 5 & 84 & 7.944-4 &         & 9.566-4 \\
  2 & 20 & 2.240+0 & 2.330+0 & 2.011+0 & 7 & 12 & 4.551-2 &         & 4.430-2 \\
  2 &192 & 1.928-1 & 1.880-1 &         & 7 & 13 & 4.547-1 & 4.580-1 & 4.490-1 \\
  3 & 21 & 3.023-2 &         & 2.279-2 & 7 & 52 & 8.013-2 & 1.500-2 & 6.685-2 \\
  3 & 38 & 4.486-3 &         & 4.188-3 & 7 & 54 & 2.256+0 & 2.120+0 & 2.136+0 \\
  3 & 44 & 1.258+0 & 1.300+0 & 1.091+0 & 8 & 12 & 7.510-3 & 7.800-1 & \\
  3 & 51 & 3.880-2 &         & 3.795-2 & 8 & 53 & 1.689-1 &         & 1.171-1 \\
  3 & 74 & 4.216-4 &         & 5.116-4 & 8 &107 & 6.846-3 &         & 4.965-3 \\
\end{tabular}
\end{ruledtabular}
\end{table*}

\setcounter{table}{5}
\begin{table*}
\caption{Comparison of weight oscillator strength ($gf$) for some
transitions of Si XI. CT01 refers to the work of Coutinho \&
Trigueiros~\cite{CT01}.}
\begin{ruledtabular}
\begin{tabular}{rrrrrrrrrr}
$i$ & $j$ & FAC & CT01 & Chianti & $i$ & $j$ & FAC & CT01 & Chianti \\
            \hline
   1  &   5  & 2.666-1  &         & 2.652-1 &   4  &   8  & 3.755-1  & 4.010-1 & 3.809-1 \\
   1  &  13  & 5.533-1  &         & 4.557-1 &   4  &  11  & 1.702-1  & 1.620-1 & 1.403-1 \\
   1  &  15  & 3.592-2  &         & 1.085-1 &   4  &  17  & 3.620-2  &         & 3.607-2 \\
   1  &  22  & 3.744-4  &         & 9.865-4 &   4  &  18  & 5.418-1  &         & 5.409-1 \\
   2  &   7  & 1.020-1  & 1.090-1 & 1.034-1 &   4  &  31  & 1.651-1  & 1.730-1 & 2.003-5 \\
   2  &  17  & 7.234-1  & 7.120-1 & 7.291-1 &   4  &  32  & 3.683-1  & 3.730-1 & 4.017-1\\
   2  &  29  & 4.403-2  &         & 5.090-2 &   4  &  73  & 1.180-1  & 1.100-1 & \\
   2  &  53  & 1.342-1  & 1.300-1 &         &   4  &  68  & 1.623-1  & 1.590-1 & \\
   3  &   7  & 7.572-2  & 8.080-2 & 7.679-2 &   4  & 107  & 2.096-1  & 1.970-1 &\\
   3  &   8  & 1.286-1  & 1.370-1 & 1.304-1 &   4  & 125  & 1.209-1  & 1.040-1 & \\
   3  &   17 & 5.423-1  &         & 5.446-1 &   5  &   9  & 2.881-1  & 3.050-1 & 2.969-1 \\
   3  &  18  & 1.625+0  & 1.600+0 & 1.634+0 &   5  &  10  & 1.913-1  & 2.340-1 & 1.922-1 \\
   3  &  27  & 2.283-1  & 2.380-1 &         &   5  &  12  & 4.144-2  &         & 3.368-2  \\
   3  &  29  & 1.166-1  & 1.200-1 & 1.187-1 &   5  &  20  & 1.638+0  & 1.670+0 & 1.482+0 \\
   3  &  32  & 9.849-2  & 1.000-1 & 1.046-1 &   5  &  37  & 7.681-1  & 7.390-1 & 8.435-1 \\
   3  &   37 & 1.226-3  &         & 1.128-3 &   5  &  44  & 8.528-2  & 8.600-1 & \\
   3  &  54  & 3.016-1  & 2.930-1 &         &   5  &  56  & 3.629-1  & 3.620-1 & \\
   3  & 106  & 1.127-1  & 1.060-1 &         &   5  &  77  & 1.622-1  & 1.600-1 & \\
   4  &   7  & 1.235-1  & 1.320-1 & 1.254-1 &   5  & 108  & 1.543-1  & 1.480-1 & \\
\end{tabular}
\end{ruledtabular}
      \end{table*}

\setcounter{table}{12}
\begin{table*}
\caption[I]{Comparison of effective collision strengths
    $\Upsilon_{ij}$ (written in form a$\pm$b implying a$\times10^{\pm b}$)
    of Si IX between present results and those used by the Chianti code.
    The data at three temperatures of 0.5, 1.0 and 6.0~MK, is listed for some transitions.
    For each transition, the first row represents present results and
    the second row represents data included in the Chianti database.}
\begin{ruledtabular}
\begin{tabular}{rrcccrrccc}
  &   & \multicolumn{3}{c}{$\Upsilon$} & &  & \multicolumn{3}{c}{$\Upsilon$} \\
i & j & 0.5 & 1.0 & 6.0 & i & j & 0.1 & 1.0 & 6.0 \\
            \hline
  1 &   2 & 8.359-2  & 7.295-2  & 3.641-2 & 2  & 3 & 2.485-1 & 2.615-1 & 1.810-1 \\
    &     & 2.433-1  & 1.755-1  & 1.289-1 &    &   & 6.984-1 & 4.981-1 & 3.428-1 \\
  1 &   3 & 6.440-2  & 7.631-2  & 6.087-2 & 2  & 4 & 1.343-1 & 1.163-1 & 5.587-2 \\
    &     & 1.761-1  & 1.242-1  & 8.295-2 &    &   & 1.773-1 & 1.515-1 & 1.275-1 \\
  1 &   4 & 4.305-2  & 3.728-2  & 1.791-2 & 2  & 5 & 1.862-2 & 1.588-2 & 7.105-3 \\
    &     & 5.912-2  & 5.051-2  & 4.236-2 &    &   & 2.285-2 & 1.969-2 & 1.836-2 \\
\end{tabular}
\end{ruledtabular}
      \end{table*}

\setcounter{table}{13}
\begin{table*}
\caption[I]{Comparison of effective collision strength
$\Upsilon_{ij}$
    for some excitations of Si X. Same as Table 13}
\begin{ruledtabular}
\begin{tabular}{rrcccrrccc}
  &   & \multicolumn{3}{c}{$\Upsilon$} & &  & \multicolumn{3}{c}{$\Upsilon$} \\
i & j & 0.5 & 1.0 & 6.0 & i & j & 0.1 & 1.0 & 6.0 \\
            \hline
  1 &   2 & 1.899-1  & 1.966-1  & 1.400-1 & 1 &122 & 2.075-4 & 2.221-4 & 3.524-4 \\
    &     & 7.373-1  & 4.944-1  & 1.661-1 &   &    & 3.660-4 & 4.046-4 & 6.604-4 \\
  1 &   4 & 2.130-2  & 1.936-2  & 1.074-2 & 2 & 5  & 5.928-2 & 5.399-2 & 3.054-2 \\
    &     & 3.578-2  & 2.788-2  & 1.466-2 &   &    & 1.011-1 & 7.918-2 & 4.192-2 \\
  1 &   6 & 2.356-2  & 2.140-2  & 1.191-2 & 2 &  7 & 1.340-1 & 1.378-1 & 1.605-1 \\
    &     & 4.998-2  & 3.643-2  & 1.737-2 &   &    & 1.577-1 & 1.382-1 & 1.013-1 \\
\end{tabular}
\end{ruledtabular}
      \end{table*}

\setcounter{table}{14}
\begin{table*}
    \caption[I]{Comparison of effective collision strength $\Upsilon_{ij}$
    for some excitations of Si XI. Same as in Table 13}
\begin{ruledtabular}
\begin{tabular}{rrcccrrccc}
  &   & \multicolumn{3}{c}{\rm $\Upsilon$}  &   & \multicolumn{3}{c}{\rm $\Upsilon$}  \\
i & j & 0.5 & 1.0 & 6.0 & i & j & 0.1 & 1.0 & 6.0 \\
            \hline
 1 &   5 & 1.103+0  & 1.140+0  & 1.493+0 & 1 & 45 & 5.110-4 & 5.471-4 & 8.120-4 \\
   &     & 1.073+0  & 1.134+0  & 1.458+0 &   &    & 2.225-4 & 2.344-4 & 3.140-4 \\
 1 &  12 & 5.331-2  & 5.378-2  & 6.001-2 & 1 & 46 & 3.189-3 & 3.426-3 & 5.611-3 \\
   &     & 5.717-2  & 5.847-2  & 6.454-2 &   &    & 5.589-3 & 5.897-3 & 8.151-3 \\
 1 &  13 & 2.371-2  & 2.799-2  & 6.662-2 & 2 &  3 & 5.363-2 & 4.938-2 & 2.486-2 \\
   &     & 2.390-2  & 2.734-2  & 5.449-2 &   &    & 1.677-1 & 1.203-1 & 4.522-2 \\
\end{tabular}
\end{ruledtabular}
      \end{table*}

\newpage
\newpage
\vspace{100cm} \tabcolsep 0.28cm \noindent
\section{EXPLANATION OF TABLES}
\begin{tabular}{lp{3.5cm}p{8cm}}
{\bf TALBE I.}  & \multicolumn{2}{l}{Energy levels (in {\it
Ryd}) of Si IX ion.}\\
          &   Index             &   A number assigned to each level                \\
          &   Configuration     &   The configuration                               \\
          &   Expt.             &   The data is from NIST website at \url{http://physics.nist.gov/cgi-bin/AtData/main_asd} \\
          &   FAC               &   Data from FAC code \\
          &   CIV3              &   The work of Aggarwal~\cite{Agg98} \\
          &   OTC99             &   The work of Orloski et al.~\cite{OTC99} \\
          &   Chianti           &   The work of Bhatia \& Doschek~\cite{BD93} \\
\end{tabular}

\vspace{20cm}
\newpage
\setcounter{table}{0}
\begin{longtable*}{llcrrrrr}
\caption{Energy levels (in {\it
Ryd}) of Si IX ion.} \\
\hline \hline
Index & Configuration & LSJ & Expt. & FAC & CIV3 & OTC99 & Chianti \\
\hline
\endfirsthead
\multicolumn{2}{l}{Table I--{\it continued} ...} \\
\hline\hline
 {\rm Index}& {\rm Configuration} & LSJ & {\rm Expt.} &{\rm FAC} & {\rm CIV3} & {\rm OTC99} & {\rm Chianti} \\
 \hline
\endhead
\hline
\endfoot
   1 & $  2s^22p^2      $ & $ ^3P_0 $&         &  0.0000 &  0.0000 &           &  0.000 \\
   2 & $  2s^22p^2      $ & $ ^3P_1 $&  0.0232 &  0.0226 &  0.0226 &           &  0.023 \\
   3 & $  2s^22p^2      $ & $ ^3P_2 $&  0.0585 &  0.0570 &  0.0575 &           &  0.058 \\
   4 & $  2s^22p^2      $ & $ ^1D_2 $&  0.4823 &  0.5081 &  0.5110 &           &  0.506 \\
   5 & $  2s^22p^2      $ & $ ^1S_0 $&  0.9823 &  0.9949 &  1.0500 &           &  0.948 \\
   6 & $  2s2p^3        $ & $ ^5S_2 $&  1.3690 &  1.3025 &  1.3416 &           &  1.261 \\
   7 & $  2s2p^3        $ & $ ^3D_3 $&  2.6631 &  2.6893 &  2.6867 &   2.6621  &  2.673 \\
   8 & $  2s2p^3        $ & $ ^3D_2 $&  2.6636 &  2.6898 &  2.6871 &   2.6627  &  2.670 \\
\hline
      \end{longtable*}

\newpage
\begin{longtable*}{llcrrrr}
\caption{} \\
\hline \hline
Index& Configuration & LSJ & Expt. & FAC & CLT00 & Chianti \\
\hline
\endfirsthead
\multicolumn{2}{l}{Table II{\it --continued} ...} \\
\hline\hline
Index & {\rm Configuration} & LSJ & Expt. & FAC & CLT00 & Chianti \\
 \hline
\endhead
\hline
\endfoot
   1 & $  2s^22p            $ & $ ^2P_{1/2}  $&         &  0.0000 &         &   0.0000  \\
   2 & $  2s^22p            $ & $ ^2P_{3/2}  $&  0.0637 &  0.0623 &  0.0636 &   0.0637  \\
   3 & $  2s2p^2            $ & $ ^4P_{1/2}  $&  1.4672 &  1.4335 &  1.4756 &   1.4672  \\
   4 & $  2s2p^2            $ & $ ^4P_{3/2}  $&  1.4898 &  1.4567 &  1.4977 &   1.4898  \\
   5 & $  2s2p^2            $ & $ ^4P_{5/2}  $&  1.5224 &  1.4880 &  1.5304 &   1.5224  \\
   6 & $  2s2p^2            $ & $ ^2D_{5/2}  $&  2.6231 &  2.6622 &  2.6230 &   2.6231  \\
   7 & $  2s2p^2            $ & $ ^2D_{3/2}  $&  2.6234 &  2.6620 &  2.6231 &   2.6234  \\
   8 & $  2s2p^2            $ & $ ^2S_{1/2}  $&  3.3505 &  3.4029 &  3.3503 &   3.3505  \\
\end{longtable*}

\newpage
\begin{longtable*}{llcrrrr}
\caption{} \\
\hline\hline
Index& Configuration & LSJ & Expt. & FAC  & CT01 & Chianti \\
 \hline
\endfirsthead
\multicolumn{2}{l}{Table III{\it --continued} ...} \\
\hline\hline
Index& Configuration & LSJ & Expt. & FAC  & CT01 & Chianti \\
 \hline
\endhead
\hline
\endfoot
   1  & $  2s^2    $ & $^1S_0  $&           &   0.000000  &   0.000000  &   0.000000   \\
   2  & $  2s2p    $ & $^3P_0  $&   1.547354&   1.544777  &   1.547362  &   1.556960   \\
   3  & $  2s2p    $ & $^3P_1  $&   1.568696&   1.565661  &   1.568859  &   1.578420   \\
   4  & $  2s2p    $ & $^3P_2  $&   1.615845&   1.611971  &   1.615744  &   1.625680   \\
   5  & $  2s2p    $ & $^1P_1  $&   3.004365&   3.050184  &   3.004353  &   3.131050   \\
   6  & $  2p^2    $ & $^3P_0  $&   4.043030&   4.031292  &   4.043251  &   4.069800   \\
   7  & $  2p^2    $ & $^3P_1  $&   4.068764&   4.060359  &   4.068830  &   4.095150   \\
   8  & $  2p^2    $ & $^3P_2  $&   4.109507&   4.099710  &   4.109646  &   4.136090   \\
\end{longtable*}

\newpage
\setcounter{table}{6}
\begin{longtable*}{llcccccccccc}
\caption{} \\
\hline\hline
i & j & \multicolumn{10}{c}{Collision strength $\Omega$}\\
 & & \multicolumn{10}{c}{Scattered electron energy (eV)} \\
 & & 10 & 50 & 100 & 200 & 400 & 600 & 800 & 1000 & 1500 & 2000 \\
 \hline
\endfirsthead
\multicolumn{6}{l}{Table VII{\it --continued} ...} \\
\hline\hline
i & j & \multicolumn{10}{c}{Collision strength $\Omega$}\\
 & & \multicolumn{10}{c}{Scattered electron energy (eV)} \\
 & & 10 & 50 & 100 & 200 & 400 & 600 & 800 & 1000 & 1500 & 2000 \\
 \hline
\endhead
\hline
\endfoot
  1 &   2 &9.040-02 &7.805-02 &6.377-02 &4.670-02 &2.733-02 &1.793-02 &1.274-02 &9.500-03 &5.314-03 &3.353-03\\
  1 &   3 &4.495-01 &2.123-01 &9.070-02 &7.488-02 &6.186-02 &3.720-02 &3.540-02 &3.593-02 &3.053-02 &2.962-02\\
  1 &   4 &3.933-02 &3.752-02 &3.260-02 &2.324-02 &1.267-02 &8.160-03 &5.565-03 &3.977-03 &2.164-03 &1.345-03\\
  1 &   5 &5.625-03 &5.129-03 &4.243-03 &2.870-03 &1.423-03 &8.418-04 &5.267-04 &3.455-04 &1.633-04 &8.897-05\\
  1 &   6 &1.274-02 &1.134-02 &9.723-03 &7.414-03 &4.620-03 &3.097-03 &2.234-03 &1.683-03 &9.516-04 &6.050-04\\
  1 &   7 &4.332-05 &3.274-05 &2.340-05 &3.693-05 &8.214-05 &3.605-05 &4.109-05 &4.763-05 &3.705-05 &3.546-05\\
  1 &   8 &1.060-02 &9.376-03 &7.984-03 &6.061-03 &3.767-03 &2.522-03 &1.819-03 &1.372-03 &7.751-04 &4.927-04\\
  1 &   9 &3.492-01 &3.815-01 &4.157-01 &4.652-01 &5.344-01 &5.842-01 &6.222-01 &6.528-01 &7.101-01 &7.532-01\\
\end{longtable*}

\newpage
\setcounter{table}{7}
\begin{longtable*}{llcccccccccc}
\caption{} \\
\hline\hline
i & j & \multicolumn{10}{c}{Collision strength $\Omega$}\\
 & & \multicolumn{10}{c}{Scattered electron energy (eV)} \\
 & & 10 & 50 & 100 & 200 & 400 & 600 & 800 & 1000 & 1500 & 2000 \\
 \hline
\endfirsthead
\multicolumn{6}{l}{Table VIII{\it --continued} ...} \\
\hline\hline
i & j & \multicolumn{10}{c}{Collision strength $\Omega$}\\
 & & \multicolumn{10}{c}{Scattered electron energy (eV)} \\
 & & 10 & 50 & 100 & 200 & 400 & 600 & 800 & 1000 & 1500 & 2000 \\
 \hline
\endhead
\hline
\endfoot
  1 &   2 &6.724-02 &2.139-01 &2.335-01 &1.958-01 &1.253-01 &9.269-02 &7.719-02 &6.914-02 &6.121-02 &5.893-02\\
  1 &   3 &1.650-02 &1.489-02 &1.309-02 &1.025-02 &6.759-03 &4.859-03 &3.722-03 &2.995-03 &2.027-03 &1.577-03\\
  1 &   4 &2.315-02 &2.080-02 &1.819-02 &1.406-02 &8.981-03 &6.205-03 &4.535-03 &3.459-03 &2.009-03 &1.318-03\\
  1 &   5 &1.388-02 &1.247-02 &1.090-02 &8.420-03 &5.373-03 &3.707-03 &2.705-03 &2.059-03 &1.188-03 &7.726-04\\
  1 &   6 &2.564-02 &2.300-02 &2.007-02 &1.555-02 &9.980-03 &6.911-03 &5.056-03 &3.859-03 &2.266-03 &1.501-03\\
  1 &   7 &6.374-01 &6.883-01 &7.419-01 &8.268-01 &9.505-01 &1.038+00 &1.104+00 &1.158+00 &1.261+00 &1.337+00\\
  1 &   8 &3.672-01 &3.939-01 &4.229-01 &4.696-01 &5.396-01 &5.904-01 &6.296-01 &6.616-01 &7.231-01 &7.686-01\\
  1 &   9 &5.617-01 &6.022-01 &6.466-01 &7.183-01 &8.268-01 &9.057-01 &9.670-01 &1.017+00 &1.113+00 &1.184+00\\
\end{longtable*}

\newpage
\setcounter{table}{8}
\begin{longtable*}{llcccccccccc}
\caption{} \\
\hline\hline
i & j & \multicolumn{10}{c}{Collision strength $\Omega$}\\
 & & \multicolumn{10}{c}{Scattered electron energy (eV)} \\
 & & 10 & 50 & 100 & 200 & 400 & 600 & 800 & 1000 & 1500 & 2000 \\
 \hline
\endfirsthead
\multicolumn{6}{l}{Table IX{\it --continued} ...} \\
\hline\hline
i & j & \multicolumn{10}{c}{Collision strength $\Omega$}\\
 & & \multicolumn{10}{c}{Scattered electron energy (eV)} \\
 & & 10 & 50 & 100 & 200 & 400 & 600 & 800 & 1000 & 1500 & 2000 \\
 \hline
\endhead
\hline
\endfoot
  1 &   2 &5.681-03 &5.114-03 &4.499-03 &3.539-03 &2.325-03 &1.634-03 &1.208-03 &9.291-04 &5.441-04 &3.558-04\\
  1 &   3 &1.742-02 &1.576-02 &1.396-02 &1.115-02 &7.611-03 &5.600-03 &4.371-03 &3.572-03 &2.485-03 &1.967-03\\
  1 &   4 &2.818-02 &2.537-02 &2.233-02 &1.756-02 &1.154-02 &8.104-03 &5.993-03 &4.609-03 &2.699-03 &1.765-03\\
  1 &   5 &1.006+00 &1.081+00 &1.162+00 &1.292+00 &1.484+00 &1.624+00 &1.731+00 &1.818+00 &1.984+00 &2.105+00\\
  1 &   6 &2.831-04 &2.423-04 &2.022-04 &1.457-04 &8.445-05 &5.488-05 &3.876-05 &2.916-05 &1.752-05 &1.277-05\\
  1 &   7 &8.171-04 &6.961-04 &5.773-04 &4.103-04 &2.293-04 &1.420-04 &9.453-05 &6.627-05 &3.202-05 &1.809-05\\
  1 &   8 &1.362-03 &1.168-03 &9.784-04 &7.116-04 &4.237-04 &2.857-04 &2.110-04 &1.669-04 &1.142-04 &9.318-05\\
  1 &   9 &6.221-03 &6.642-03 &7.076-03 &7.748-03 &8.632-03 &9.182-03 &9.550-03 &9.814-03 &1.023-02 &1.045-02\\
\end{longtable*}

\newpage
\setcounter{table}{9}
\begin{longtable*}{llccccccccc}
\caption{} \\
\hline\hline
i& j & $gf$  & $A$ & \multicolumn{7}{c}{Effective Collision strength $\Upsilon$}\\
 & & & s$^{-1}$ & 0.5 & 1.0 & 2.0 & 3.0 & 4.0 & 5.0 & 6.0 \\
 \hline
\endfirsthead
\multicolumn{6}{l}{Table X{\it --continued} ...} \\
\hline\hline
i& j & $gf$  & $A$ & \multicolumn{7}{c}{Effective Collision strength $\Upsilon$}\\
 & & & s$^{-1}$ & 0.5 & 1.0 & 2.0 & 3.0 & 4.0 & 5.0 & 6.0 \\
 \hline
\endhead
\hline
\endfoot
  1 &   2 &1.995-07 &2.721-01 &8.359-02 &7.295-02 &5.955-02 &5.094-02 &4.479-02 &4.011-02 &3.641-02\\
  1 &   3 &8.784-13 &4.586-06 &6.440-02 &7.631-02 &7.685-02 &7.239-02 &6.792-02 &6.408-02 &6.087-02\\
  1 &   4 &1.980-12 &8.212-04 &4.305-02 &3.728-02 &3.007-02 &2.550-02 &2.226-02 &1.982-02 &1.791-02\\
  1 &   5 &         &         &5.828-03 &4.970-03 &3.920-03 &3.270-03 &2.818-03 &2.483-03 &2.223-03\\
  1 &   6 &1.205-11 &3.285-02 &1.174-02 &1.055-02 &8.875-03 &7.722-03 &6.867-03 &6.201-03 &5.665-03\\
  1 &   7 &         &         &2.940-05 &3.164-05 &3.942-05 &4.347-05 &4.522-05 &4.590-05 &4.608-05\\
  1 &   8 &8.687-11 &1.010+00 &9.668-03 &8.677-03 &7.289-03 &6.336-03 &5.631-03 &5.082-03 &4.641-03\\
  1 &   9 &8.160-02 &1.582+09 &3.788-01 &4.022-01 &4.390-01 &4.676-01 &4.913-01 &5.115-01 &5.292-01\\
\end{longtable*}

\newpage
\setcounter{table}{10}
\begin{longtable*}{llccccccccc}
\caption{} \\
\hline\hline
i& j & $gf$  & $A$ & \multicolumn{7}{c}{Effective Collision strength $\Upsilon$}\\
 & & & s$^{-1}$ & 0.5 & 1.0 & 2.0 & 3.0 & 4.0 & 5.0 & 6.0 \\
 \hline
\endfirsthead
\multicolumn{6}{l}{Table XI{\it --continued} ...} \\
\hline\hline
i& j & $gf$  & $A$ & \multicolumn{7}{c}{Effective Collision strength $\Upsilon$}\\
 & & & s$^{-1}$ & 0.5 & 1.0 & 2.0 & 3.0 & 4.0 & 5.0 & 6.0 \\
 \hline
\endhead
\hline
\endfoot
  1 &   2 &3.683-07 &2.871+00 &1.899-01 &1.966-01 &1.862-01 &1.717-01 &1.591-01 &1.486-01 &1.400-01\\
  1 &   3 &3.261-05 &2.691+05 &1.523-02 &1.389-02 &1.192-02 &1.052-02 &9.472-03 &8.649-03 &7.982-03\\
  1 &   4 &1.404-06 &5.983+03 &2.130-02 &1.936-02 &1.648-02 &1.445-02 &1.292-02 &1.172-02 &1.074-02\\
  1 &   5 &2.090-11 &6.194-02 &1.277-02 &1.160-02 &9.875-03 &8.655-03 &7.737-03 &7.015-03 &6.430-03\\
  1 &   6 &2.173-10 &2.061+00 &2.356-02 &2.140-02 &1.824-02 &1.600-02 &1.431-02 &1.299-02 &1.191-02\\
  1 &   7 &1.421-01 &2.022+09 &6.756-01 &7.173-01 &7.821-01 &8.325-01 &8.741-01 &9.097-01 &9.409-01\\
  1 &   8 &1.126-01 &5.238+09 &3.874-01 &4.101-01 &4.459-01 &4.743-01 &4.980-01 &5.184-01 &5.364-01\\
  1 &   9 &1.900-01 &1.007+10 &5.924-01 &6.271-01 &6.822-01 &7.261-01 &7.628-01 &7.944-01 &8.224-01\\
\end{longtable*}

\newpage
\setcounter{table}{11}
\begin{longtable*}{llccccccccc}
\caption{} \\
\hline\hline
i& j & $gf$  & $A$ & \multicolumn{7}{c}{Effective Collision strength $\Upsilon$}\\
 & & & s$^{-1}$ & 0.5 & 1.0 & 2.0 & 3.0 & 4.0 & 5.0 & 6.0 \\
 \hline
\endfirsthead
\multicolumn{6}{l}{Table XII{\it --continued} ...} \\
\hline\hline
i& j & $gf$  & $A$ & \multicolumn{7}{c}{Effective Collision strength $\Upsilon$}\\
 & & & s$^{-1}$ & 0.5 & 1.0 & 2.0 & 3.0 & 4.0 & 5.0 & 6.0 \\
 \hline
\endhead
\hline
\endfoot
  1 &   2 &         &         &5.049-03 &4.672-03 &4.154-03 &3.669-03 &3.192-03 &2.821-03 &2.525-03\\
  1 &   3 &4.835-05 &3.173+05 &1.558-02 &1.447-02 &1.295-02 &1.154-02 &1.015-02 &9.068-03 &8.208-03\\
  1 &   4 &5.669-11 &2.367-01 &2.505-02 &2.318-02 &2.061-02 &1.820-02 &1.584-02 &1.400-02 &1.253-02\\
  1 &   5 &2.666-01 &6.641+09 &1.103+00 &1.140+00 &1.219+00 &1.295+00 &1.373+00 &1.438+00 &1.493+00\\
  1 &   6 &         &         &2.327-04 &2.131-04 &1.877-04 &1.631-04 &1.355-04 &1.160-04 &1.008-04\\
  1 &   7 &4.846-10 &2.139+01 &6.676-04 &6.098-04 &5.347-04 &4.620-04 &3.804-04 &3.227-04 &2.776-04\\
  1 &   8 &6.542-10 &1.767+01 &1.123-03 &1.031-03 &9.108-04 &7.951-04 &6.650-04 &5.733-04 &5.018-04\\
  1 &   9 &1.497-07 &4.958+03 &6.800-03 &6.963-03 &7.319-03 &7.662-03 &8.028-03 &8.307-03 &8.536-03\\
\end{longtable*}

\begin{references}
\bibitem{RMA02}
      A.J.J. Raassen, R. Mewe, M. Audard, et al., Astron. Astrophys. 389 (2002) 228
\bibitem{RNM03}
      A.J.J. Raassen, J.-U. Ness, R. Mewe, et al., Astron. Astrophys. 400 (2003) 671
\bibitem{ABG01}
      M. Audard, E. Behar, M. G${\rm \ddot{u}}$del, et al., Astron. Astrophys. 365 (2001) L329
\bibitem{AKN05}
      K.M. Aggarwal, F.P. Keenan, S. Nakazaki, Astron. Astrophys. 436 (2005) 1141
\bibitem{LDZ04}
      G.Y. Liang, G.X. Dong, J. L. Zeng, At. Data Nucl. Data Tables 88 (2004) 83
\bibitem{LB06}
      E. Landi, A.K. Bhatia, At. Data Nucl. Data Tables 92 (2006)
      305
\bibitem{LZJ06}
      G.Y. Liang, J.Y. Zhong, Z. Jin, et al., (2006) (in preparation)
\bibitem{Gu03}
      M.F. Gu, Astronphys. J. 582 (2003) 1241
\bibitem{Gu05}
      M.F. Gu, Astronphys. J. Suppl. 156 (2005) 105
\bibitem{DGJ89}
      K. G. Dyall, I. P. Grant, C. T. Johnson, et al., Comput. Phys. Commun. 55 (1989) 424
\bibitem{Agg98}
      K.M. Aggarwal, Astrophys. J. Supp. 118 (1998) 589
\bibitem{BD93}
      A.K. Bhatia, G.A. Doschek, At. Data Nucl. Data Tables 55 (1993) 281
\bibitem{OTC99}
      R.V. Orloski, A.G. Trigueiros, G.H. Cavalcanti, J.
      Quant. Spectrosc. Radiat. Transfer 61 (1999) 665
\bibitem{RK69}
      L.J. Radziemski, V. Kaufman, J. Opt. Soc. Am. 59 (1969) 424
\bibitem{Edl83}
      B. Edlen, Phys. Scripta 28 (1983) 483
\bibitem{ZS94}
      H.L. Zhang, D.H. Sampson, At. Data Nucl. Data Tables 58 (1994)
      255
\bibitem{ZGP94}
      H.L. Zhang, M. Graziani, A.K. Prahan, Astron. Astrophys. 283
      (1994) 319
\bibitem{SZ95}
      D.H. Sampson, H.L. Zhang, unpublished calculation (1995)
\bibitem{CLT00}
      G.H. Cavalcanti, F.R.T. Luna, A.G. Trigueiros, J.
      Quant. Spectrosc. Radiat. Transfer 64 (2000) 5
\bibitem{SG84}
      D.H. Sampson, S.J. Goett, R.E.H. Clark, At. Data Nucl. Data Tables
      30 (1984) 125
\bibitem{ZS92}
      H.L. Zhang, D.H. Sampson, At. Data Nucl. Data Tables 52 (1992) 143
\bibitem{CT01}
      L. H. Coutinho, A.G. Trigueiros, J.
      Quant. Spectrosc. Radiat. Transfer 68 (2001) 643
\bibitem{Cow81}
      R. D. Cowan, The theory of atomic structure and spectra, Berkeley, CA: University of California
      Press (1981)
\bibitem{SZM89}
      D.H. Sampson, H.L. Zhang, A. K. Mohanty, R.E.H. Clark,
      Phys. Rev. A 40 (1989) 604
\bibitem{Agg83} K.M. Aggarwal, K.L. Baluja J. Phys. B 16 (1983) 107
\bibitem{Hib75}
      A. Hibbert, Comput. Phys. Commun. 9 (1975) 141
\bibitem{MB78}
      H.E. Mason, A.K. Bhatia, Mon. Not. Royal Astron. Soc. 184 (1978) 423
\bibitem{ZS96}
      H.L. Zhang, D.H. Sampson, At. Data Nucl. Data Tables 63 (1996) 275
\bibitem{KOT00}
      F.P. Keenan, E. O$^,$Shea, R.J. Thomas, et al., Mon. Not. Royal
      Astron. Soc. 315 (2000) 450
\bibitem{GAB01}
      M. G${\rm \ddot{u}}$del, M. Audard, K. Briggs, et al., Astron. Astrophys. 365 (2001) L336
\bibitem{BBD85}
      K.A. Berrington, P.G. Burke, P.L. Dufton, A.E. Kingston, At. Data Nucl. Data Tables, 33 (1985) 195
\end{references}
\end{document}